\documentclass[preprint]{aastex}
\begin{document}
\begin{center} \large{\bf Luminosity profiles of advanced mergers of galaxies
using 2MASS data\\}
\end{center}
\noindent
\centerline{Aparna Chitre and Chanda J.Jog}
\noindent
\centerline{Department of Physics, Indian Institute of Science, Bangalore 560 012, INDIA}
\vskip 4in
\noindent
Running title : Luminosity profiles of advanced mergers\\
\newpage
{\bf Abstract}
A sample of 27 disturbed galaxies
that show signs of
interaction but have a single nucleus were selected from the Arp and the
Arp-Madore catalogues.
For these, the K$_s$ band images from 
the Two Micron All Sky Survey (2MASS) are analysed to obtain their radial
luminosity profiles and other structural parameters.
We find that in spite of their similar optical
appearance, the sample galaxies vary in their dynamical properties
, and fall
into two distinct classes. The first class consists of galaxies which can
be described by a single $r^{1/4}$ law and the second class consists
of galaxies
that show an outer exponential disk.
A few galaxies that have disturbed
profiles cannot be fit into either of the above classes. However,
all the galaxies are similar in all other parameters such as the far-infrared
colours, the molecular hydrogen content and the central velocity
dispersion. Thus, the dynamical parameters of these sets seem to be determined
by the ratio of the initial masses of the colliding galaxies.
We propose that the galaxies in the first class result from a merger
of spiral galaxies of equal
masses whereas the second class of galaxies results from a merger
of unequal mass
galaxies. The few objects that do not fall into either category show a
disturbed luminosity profile and a wandering centre,
which is indicative of these being
unrelaxed mergers.
Of the 27 galaxies in our
sample, 9 show elliptical-like profiles and 13 show an outer
exponential. Interestingly, Arp 224, the second oldest merger remnant
of the Toomre sequence shows an exponential disk in the outer parts.\\
{\em keywords:}Galaxies:kinematics and dynamics -Galaxies:evolution -Galaxies:interactions -Galaxies:photometry -Galaxies:structure

\section{Introduction}
It was proposed theoretically a long time ago that a
merger of a pair of equal-mass spirals gives rise to an elliptical galaxy
(Toomre 1977).
This has been confirmed observationally by near-IR studies of a few
mergers which indeed show an
$r^{1/4}$ de Vaucouleur's radial profile typical of ellipticals
(Schweizer 1982, Joseph \& Wright 1985,
Wright et al. 1990, Stanford \& Bushouse 1991
(SB91 hereafter)). Theoretical
work, involving mergers of galaxies (simulated using stars alone), has shown that under
some conditions mergers indeed give rise to ellipticals with the above
radial profile (Barnes \& Hernquist 1991, Hernquist 1992).
The mergers simulated with star formation in gas-rich spirals (Bekki 1998a)
also give a similar profile. 

Despite this overall agreement, there are many issues
which need to be answered. First, the origin
of the $r^{1/4}$ profile is not physically well understood, although
violent relaxation is believed to be responsible for
the elliptical light distribution (Lynden-Bell 1967).
Both collisionless relaxation in an isolated galaxy (van Albada 1982)
as well as in a merger (e.g. Barnes 1988) are known to result in an
$r^{1/4}$ profile.
 Second, it is not clear if an $r^{1/4}$
profile over the entire galaxy is the
only outcome that is possible for a galaxy merger, because the
parameter space for galaxy interaction has not yet been
studied exhaustively by the N-body theoretical work in the literature.
The real galaxy mergers
will presumably tell us about the parameter space that needs to be
explored in future theoretical work.
Third, other properties such as asymmetry in morphology, twisted isophotes,
boxy/disky isophotes, shells, etc. have been shown to be good indicators
that an elliptical galaxy was formed by a merger event (Nieto \& Bender 1989; Hernquist \& Spergel 1992).

In order to shed light on some of these issues, in this paper we have analyzed
the near infrared K$_s$ band images from the 2MASS database for a sample of
galaxies that show evidence of an advanced merger and we have also deduced other
structural parameters.
The large size of the database allows us to construct a big sample.
The K$_s$ band best traces the underlying mass distribution in a galaxy. Also,
the extinction in the K$_s$ band is very low, which enables us to study the
underlying stellar distribution free of the effects of dust obscuration
and young star-forming regions. We have chosen a
sample of colliding galaxies from the Arp (1966) and Arp-Madore (1987)
catalogs based purely on their optical appearance e.g. galaxies that
show strong signs of
interaction such as tails and a distorted, puffed-up main body
but have a single merged nucleus.
A similar study was first done by
SB91 for ten galaxies, except that they considered the
profiles only along the major axis and the minor axis. The
subsequent papers to look at this
problem have chosen their samples from IRAS (James et al. 1999,
Scoville et al. 2000, Zheng et al. 1999) and are therefore biased
towards high-luminosity,
 distant galaxies. Our sample, on the other hand, is unbiased with respect to the
IR luminosity of a galaxy.

In this paper, we extend the  study of SB91 for a much
larger sample of interacting galaxies, and study their radial
profiles as well as other structural parameters.
We give new detections of the $r^{1/4}$ profile in four galaxies
from the Arp catalogue
and thus our work has extended the known number of mergers showing such profiles.
We have looked at the full azimuthal data and thus the luminosity of
the galaxies is sampled in a better way.
We check what fraction of our sample shows the
 $r^{1/4}$ profile expected for an elliptical galaxy and
compare our results with those of the
SB91, and other previous work. We also study other
structural parameters such as
boxiness/diskiness of isophotes, wandering centres etc. that provide
important additional information on mergers.
         In Section 2, we describe the criteria for selection
for our sample and also the data acquired from the public sites.
In Section 3, we describe the data analysis. In
Section 4, we discuss the results and our conclusions are
summarized in Section 5.

\begin{table*}
\caption[]{The sample}
\label{sample}
\begin{tabular}{clllcc}
\hline
Galaxy&Alt.name&RA(2000)&Dec(2000)&Type(RC3)&Type(SIMBAD)\\
\hline
&&&&&\\
Arp156&UGC5814&10:42:38.0&+77:29:41&-&-\\
Arp160&NGC4194&12:14:09.8&+54:31:39&IBm pec&I\\
Arp162&NGC3414 &0:51:16.3&+27:58:33&S0 pec&SB0\\
Arp163&NGC4670&12:45:17.0&+27:07:34&SB(s)0/a pec&S0\\
Arp165&NGC2418&07:36:37.9&+17:53:06&E&E\\
Arp186&NGC1614&04:33:59.9&-08:34:30&SB(s)c pec&I\\
Arp187&-&05:04:52.99&-10:14:51&-&-\\
Arp193&IC883 &13:20:35.3&+34:08:22&Im: pec&-\\
Arp212&NGC7625&23:20:30.8&+17:13:41&SA(rs)a pec&Sp\\
Arp214&NGC3718&11:32:35.7&+53:03:59&SB(s)a pec&SB0/Sa\\
Arp219&-&03:39:53.26&-02:06:47&-&SB+cG\\
Arp221&-&09:36:28.03&-11:19:49&-&-\\
Arp222&NGC7727&23:39:53.8&-12:17:36&SAB(s)a pec&Sa\\
Arp224&NGC3921&11:51:06.1&+55:04:39&(R$^\prime$)SA(s)0/a pec&S0\\
Arp225&NGC2655&08:55:38.7&+78:13:28&SAB(s)0/a&S0/Sa\\
Arp226&NGC7252 &22:20:44.9&-24:40:41&(R)SA(r)&-\\
Arp230&IC51&00:46:24.3&-13:26:34&S0 pec?&-\\
Arp231&PGC2616&00:43:54.3&-04:14:43&-&S0+\\
Arp243&NGC2623&08:38:24.2&+25:45:01&-&-\\
AM0318-230&PGC12526&03:20:40.1&-22:55:50&S?&Sa\\
AM0337-312&NGC1406&03:39:22.6&-31:19:19&SB(s)bc: sp&Sc\\
AM1315-263&NGC5061 &13:18:04.9&-26:50:10&E0&E\\
AM2146-350&NGC7135&21:49:43.7&-34:52:42&SA0- pec&S\\
AM0501-632&PGC16567&05:01:30.1&-63:17:34&S0+? pec&S\\
AM1300-233&MGC-4-31-23&13:02:52.07 &-23:55:19&-&Ir\\
AM1306-282&PGC45596&13:09:10.4&-28:38:23&SAB(rs)s pec?&Sa\\
AM1324-431&PGC47188 &13:27:51.0&-43:25:50&S?&Sc\\
\end{tabular}
\end{table*}

\begin{table*}
\caption[]{Basic data}
\label{bdata}
\begin{tabular}{clllll}
\hline
Galaxy&f$_{100}/$f$_{60}$&L$_{IR}$&M$_B$&radius\arcsec\ &K$_s$\\
& &($\times$ $10^{10} L_{\odot}$)&&2MASS&2MASS\\
\hline
&&&&&\\
Arp 156&2.61&$<$0.21&-17.142&26.8&10.534\\
Arp 160&1.21&7.62&-20.322&25.6&9.680\\
%Arp 162&$<$4.8&0.008$<$L$<$0.078&-19.606&64.4&8.059\\
Arp 162&$<$4.8&$<$0.07&-19.606&64.4&8.059\\
%Arp 163&1.69&0.18$<$L$<$0.217&-17.9&30&10.612\\
Arp 163&1.69&$<$0.21&-17.9&30&10.612\\
Arp 165&-&$<$0.27&-20.76&40.6&9.037\\
%Arp 168&-&$<$0.004&-17.576&99&5.245\\
%Arp 185&1.169&-20.069&80\\
Arp 186&1.01&39.4&-21.193&24.2&9.565\\
Arp 187&-&-&-&16.8&11.314\\
Arp 193&1.63&36.9&-20.38&19.6&10.895\\
Arp 212&1.90&1.45&-19.081&42.8&8.929\\
Arp 214&3.28&0.07&-19.818&80&7.852\\
%Arp 219&2.02&3.9$<$L$<$5.69&-21.411&19.2&11.462\\
Arp 219&2.02&$<$5.69&-21.411&19.2&11.462\\
Arp 221&1.63&-&-20.376&33.8&9.491\\
Arp 222&-&-&-20.546&68.6&7.797\\
%Arp 224&$<$2.45&0.94$<$L$<$2.42&-21.58&27.8&9.939\\
Arp 224&$<$2.45&$<$2.42&-21.58&27.8&9.939\\
Arp 225&2.98&0.25&-20.803&80&7.223\\
Arp 226&1.76&5.31&-20.926&33.2&9.399\\
Arp 230&2.12&0.41&-18.089&25.6&10.364\\
Arp 231&-&-&-19.713&26.4&9.887\\
Arp 243&1.20&29.33&-21.205&19.8&10.505\\
AM0318-230&-&-&-20.568&17.2&11.166\\
AM0337-312&2.26&0.82&-18.506&80&8.823\\
AM1315-263&-&$<$0.07&-21.118&71.8&7.372\\
%AM2146-350&3.5&0.09$<$L$<$0.201&-19.579&51&8.998\\
AM2146-350&3.5&$<$0.20&-19.579&51&8.998\\
AM0501-632&2.60&0.06&-17.111&24.2&10.851\\
AM1300-233&1.11&23.61&-21.282&37.4&10.281\\
%AM1306-282&2.05&0.06$<$L$<$0.125&-18.372&23.4&11.291\\
AM1306-282&2.05&$<$0.12&-18.372&23.4&11.291\\
%AM1324-431&1.31&9.29$<$L$<$10.74&-21.293&18&12.239\\
AM1324-431&1.31&$<$10.74&-21.293&18&12.239\\
%AM2213-370&12.44&-21.421&68.6\\
\end{tabular}
%\begin{list}{}{}
%\item Col 1 :
%\end{list}
\end{table*}

\begin{table*}
\caption{Properties of the sample galaxies}
\label{props}
\begin{tabular}{cccccccl}
\hline
Galaxy&D&1\arcsec & R$_e$$,$R$_d$&$\mu$&C$_{31}$&$\sigma$&Spectral type\\
&(Mpc)&(pc)&(kpc)&(mag arcsec$^{-2}$)&&(km sec$^{-1}$)&\\
\hline
&&&&&&&\\
Class I&&&R$_e$&$\mu_e$&&&\\
&&&&&&&\\
Arp 156&25&121&1.03$\pm$ 0.14&18.08&4.35&-&-\\
Arp 165&67&325&5.92$\pm$ 0.76&18.37&4.98&268&-early type \\
Arp 193&93&451&3.89$\pm$ 0.38&17.89&3.68&206&LINER, starburst\\
Arp 221&-&-&-&18.32&4.18&-&-\\
Arp 222&25&125&2.83$\pm$ 0.46&17.43&5.23&200&-\\
Arp 225&18&91&3.26$\pm$ 0.38&17.70&4.86&164&LINER,early type\\
Arp 226&64&312&3.28$\pm$ 0.77&17.43&4.74&177&-\\
Arp 231&76&368&3.80$\pm$ 0.45&17.77&4.13&-&-\\
Arp 243&72&351&1.58$\pm$ 0.36&16.68&4.07&95&LINER, starburst\\
&&&&&&&\\
Class II&&&R$_d$&$\mu_0$&&&\\
&&&&&&&\\
Arp 162&17&87&1.56$\pm$ 0.04&16.37&5.50&250&LINER\\
Arp 186&63&308&1.69$\pm$ 0.09&15.33&4.62&150?&AGN\\
Arp 187&164&795&4.05$\pm$0.19&16.44&3.24&-&-\\
Arp 212&21&104&0.85$\pm$ 0.03&15.26&3.33&-&starburst\\
Arp 214&13&65&1.72$\pm$0.11&16.96&5.41&178&LINER\\
Arp 219&139&673&4.86$\pm$ 0.22&17.01&3.47&-&-\\
Arp 224&79&383&2.91$\pm$ 0.21&16.57&5.14&195&post starburst\\
Arp 230&23&113&0.75$\pm$ 0.02&16.21&3.17&-&-\\
AM0318-230&142&692&4.76$\pm$ 0.51 &17.34&3.88&-&-\\
AM0337-312&14&69&2.28$\pm$ 0.07&16.21&3.33&-&-\\
AM1324-431&139&675&4.39$\pm$0.34&17.60&3.83&-&-\\
AM1315-263&24&116&1.90$\pm$ 0.04&15.65&5.66&194&-\\
AM2146-350&36&174&3.22$\pm$ 0.23&17.17&5.22&-&LINER\\
&&&&&&&\\
No fit&&&&&&&\\
&&&&&&&\\
AM0501-632&14&69&0.41$\pm$0.003&16.19&2.69&-&-\\
Arp 160&34&168&-&-&3.21&104&starburst,AGN\\
Arp 163&16&78&-&-&3.52&-&Wolf Rayet, HII\\
AM1300-233&86&417&-&-&4.47&-&Seyfert\\
AM1306-282&19&92&-&-&2.63&-&-\\
&&&&&&&\\
\end{tabular}
\end{table*}

\section{Sample selection}
The sample galaxies were chosen from the Atlas of Peculiar Galaxies
(Arp 1966) based solely on their optical appearance. Galaxies
which were classified as "neither E nor S" were chosen in order to
select merging systems that had already lost their individual
identities but had not yet reached the final smooth stage. This
class of objects from the Arp catalogue is made up of peculiar,
disturbed galaxies
that show various features such as adjacent loops or irregular clumps,
appearance of fission, etc. The Arp-Madore (1987) catalogue of Southern
Peculiar Galaxies and Associations was also used. Galaxies lying
within the categories 7 (galaxies with jets), 15 (galaxies with
tails, loops of material or debris and category 16 (irregular or
disturbed galaxies)  were selected and only those that had 2MASS data
were added to our sample.
The sample was unbiased towards any
other properties of the galaxies
such as luminosity or spectral type. The 2MASS database
at {\it http://irsa.ipac.caltech.edu} was searched and all selected galaxies
present in the 2MASS extended source
full resolution image server were chosen.
Galaxies that showed either the presence of a double nucleus or had a
diffuse appearance or poorly defined centre in the 2MASS
images were rejected. Galaxies with a radius of less then 10\arcsec
were also not considered.
The final sample consists of 27 galaxies. This sample
is
presented in the form of contour plots in Fig. 1.
Table~\ref{sample} gives the sample galaxies and the columns are as follows:
{\it Col(1)} : the name of the galaxy; {\it Col(2)} : the alternate name of the galaxy; {\it Col(3)} and {\it Col(4)} : The RA and Dec of the galaxy;
{\it Col(5)} : The type of the galaxy from RC3 and {\it Col(6)} : the type of the galaxy taken from SIMBAD.

Some basic data derived from the literature for the sample galaxies are given in Table~\ref{bdata} where:
{\it Col(1)} : the name of the galaxy;
{\it Col(2)} : the ratio of the flux at
100$\mu$ to the flux at 60$\mu$; {\it Col(3)} : The infrared luminosity between
(8-1000 $\mu$) computed using :
\begin{equation}
L_{IR} =
4 \pi D^{2} F_{IR}
\end{equation}
where
\begin{equation}
F_{IR} =
1.8 \times 10^{-14} (13.48 f_{12} + 5.16 f_{25} + 2.58 f_{60} + f_{100})
\end{equation}
from Sanders \& Mirabel (1996); {\it Col(4)} : The absolute blue
luminosity from the LEDA database ({\it http://leda.univ-lyon1.fr/}); {\it Col(5)} : the semi-major axis of
the elliptical isophote at
20 $mag\ arcsec^{-2}$ taken from the
2MASS database and {\it Col(6)} : The K$_s$ magnitude contained within
the semi-major axis in Col(5).

\section{Analysis}
The 2MASS full-resolution images for
extended sources are 3-dimensional FITS image cubes. Each cube
contains three image planes, and each image plane
contains the J,H and K$_s$ 2MASS images. The image size is scaled to
the size of the object, ranging from 21\arcsec\ $\times$ 21\arcsec\ to
101\arcsec\ $\times$ 101\arcsec\ .The 3-D FITS image cubes containing
the J,H and K$_s$ 2MASS image data were
downloaded and the K$_s$ band images were extracted from them.
The K$_s$ band images were analysed using the task ELLIPSE within STSDAS \footnote
{The Space Telescope Science Data Analysis System STSDAS is distributed by
the Space Telescope Science Institute.}. The procedure consisted of
fitting elliptical isophotes to the K$_s$ band galaxy images and deriving
the 1-dimensional azimuthally averaged radial profiles for
the surface brightness, ellipticity, position angle, etc. based on
the algorithm given by Jedrejewski (1987). The centre, position angle
and ellipticity were allowed to vary.
Ellipses were fit right up to the central pixel. The parameters thus derived,
namely the surface brightness profile, the ellipticity and the position angle
variation, the coefficient (B4) of the $cos(4\theta)$ component and the x and y coordinates of the centres were
studied.
Galaxies having a radius larger than 50\arcsec\ could not be studied beyond
that radius due to the constraint imposed by the maximum size of
101\arcsec\ $\times$ 101\arcsec\ on the full resolution images.
\subsection{Fits to the surface brightness profiles}
The 1-D surface brightness profiles obtained using the ellipse fitting
technique were examined. In each case, the magnitude zero points were taken from the FITS header
and applied to the profiles derived from the isophotal analysis to
compute the calibrated surface brightnesses. The first step was to fit the surface brightness
profiles using de Vaucouleur's law. The fits were made using the task nfit1d within STSDAS.
The fit was tried between the radius
of 2\arcsec\ to the radius where the signal falls to 1$\sigma$ of the
background noise level.
It was found that 9 of the 27 sample objects could be fit by an $r^{1/4}$ law
over this entire range. The surface brightness profiles and the fits for these
9 galaxies are
shown in Fig. 2. The profiles of 13 of the remaining galaxies
clearly show the presence of an outer exponential disk. We fit this outer
exponential and the fits are shown in Fig. 3.
The error
bars in both these figures are small except in the outer regions in
some cases. The resulting
scale lengths for the two classes are given in Table~\ref{props}.
The columns in Table~\ref{props} are:
 {\it Col(1)} :
the name of the galaxy; {\it Col(2)}: The distance to the galaxy in Mpc;
{\it Col(3)}: the linear resolution of the galaxy
 in pc per arcsec; {\it Col(4)} : the scale length (the bulge scale length
 in class I and the disk scale length in class II) in kpc; {\it Col(5)} :
 the surface brightness;
{\it Col(6)} : the concentration index C$_{31}$; {\it Col(7)} :
 the central velocity
dispersion from HYPERCAT ({\it http://www-obs.univ-lyon1.fr/hypercat/}) and {\it
Col(8)} : The spectral type for the central region taken from
SIMBAD and ADS.
Of the remaining
5 galaxies, one (AM0501-632) could be fit by a single exponential
throughout, while the other 4 (Arp 160, Arp 163, AM1300-233
and AM1306-282) could not be fit by either law. These galaxies show a
highly disturbed surface brightness profile
(Fig. 4).

\subsection{Luminosity profile classes}
Based on the behaviour of the surface brightness profiles, we classify the
sample objects into two main classes:

Class I: The surface brightness profile of these galaxies is well fit by an
$r^{1/4}$ law throughout most of its range.
The fits to Arp 165, Arp 221, Arp 222 and Arp 225 are very good right
up to the outermost data point, as shown in Fig. 2. These systems have presumably relaxed completely,
resulting in elliptical-like profiles.
A few cases, like Arp 156, Arp 193, Arp 226, Arp 231 and Arp 243, show slight
deviations in the outermost regions, which
may not have relaxed completely. Alternatively, the small deviations in the
outermost regions 
could just be the typical behaviour shown by normal ellipticals at larger radii (Binney \& Merrifield 1998).

Class II: The surface brightness profiles of this class of galaxies could not
be fit solely by a single $r^{1/4}$. When the fit was tried, we
found that there was an excess over the fit in the outer regions. A bulge
plus disk also was tried and it
does not give a good fit. In the case of Arp 212 and AM 0318-230, even
the inner region could not be fit by an $r^{1/4}$ law.
In general, we found that the combined fit was highly sensitive to the
range over which the fit was made. All these galaxies are best fit by an
exponential disk in the outer parts as shown in Fig. 3.
It should be noted that this exponential typically
fits very well over more than 2 scale lengths.

Four of the sample galaxies could not be fitted by an $r^{1/4}$ or an exponential
profile.
These systems may not
yet have fully relaxed, as suggested by their
structural parameters, like the wandering
centre (See Section 3.3). The surface brightness profiles of these galaxies are
shown in Fig. 4.

Of the 27 galaxies in our sample, 9 belong to Class I, 13 belong to Class II and 4 do not belong to either class. The galaxies Arp 222 and Arp 225 from
Class I have radii greater than 50\arcsec\ and the possible presence of an outer exponential disk beyond 50\arcsec\ cannot be ruled out completely.
However, the $r^{1/4}$ law gives a good fit up to the outermost point as is
seen in Fig. 2. We cannot make a clear statistical
remark about the percentages belonging to each class since our sample is
not a complete one.

Information on the nature of the luminosity profiles of some of
our sample galaxies exists in the literature and a brief description is
given below:
Arp 162, Arp 214, AM 0337-312 and Arp 163 show an inner truncated exponential
disk in the photographic V band and Arp 225 shows only an $r^{1/4}$ profile
(Baggett et al. 1998). Arp 212 shows an
outer exponential in the B band (Lu 1998).
Scoville et al.(2000) find that the brightness profile
of Arp 243 in the H and K bands can be
approximated by an $r^{1/4}$ profile and the H band profile of Arp 193 is
better fit by an $r^{1/4}$ law than an exponential law, based on HST data.
For Arp 224, we find that the outer part can be fit quite well
by an exponential profile as is clear from the small error bars in Fig. 3.
SB91 report that the K band surface brightness profile can be fit by an
$r^{1/4}$ law only in the inner 8\arcsec\  and the profile deviates from
an $r^{1/4}$ law beyond this point.
Schweizer(1996) finds no evidence for the presence of an
outer exponential disk in the R band images of Arp 224.

\subsection{Structural Parameters} 
In addition to the luminosity profile, the results of the isophotal analysis
also provide valuable information about other structural and
the dynamical parameters of the galaxies.
The coefficient (B4) of the
$cos(4\theta)$ component is a measure of the diskiness (if B4 $>$ 0)
or boxiness (if B4 $<$ 0) of the
galaxy. Studies of elliptical galaxies show that boxiness is an indicator
of a merger as the origin of some ellipticals (Nieto \& Bender 1989).
Our sample however shows a disturbed morphology, yet is fit by an $r^{1/4}$
law (Class I). Therefore, it is not clear what physical significance the values of B4
have for these disturbed systems. Nevertheless, we give these trends in
Appendix A for each galaxy, which may be used as input for future theory work by
others. The analysis also provides information about the shift in the
centre for successive isophotes.
The x-y coordinates of the centre of each fitted isophote can tell us
whether the isophote is concentric or otherwise. 
If the isophotes are not concentric, it means that there is asymmetry in the
central mass distribution.

Details on each object can be found in Appendix A.
Of the 27 sample galaxies, we find that
10 show only diskiness, 4 show only boxiness while 8 show both boxiness and
diskiness. 5 galaxies do not show any clear trend towards diskiness or boxiness.
For the classes I and II, we find that the inner isophotes are concentric.
The coordinates of the centre start changing for the outer isophotes.
In contrast, for all the unrelaxed objects shown in Fig. 4,
the isophotes are non-concentric right from the innermost regions. That is,
the centres of the isophotes show a wandering or a "sloshing" pattern, indicating
the dynamically unrelaxed behaviour of the merger {\it even in the central regions}. This difference is clearly illustrated in Fig. 5 where
the quantities derived
from the isophotal analysis , namely A4, B4, X$_0$ and Y$_0$ are
presented for one galaxy belonging to Class I (Arp 193) and
one which could not be fit by an $r^{1/4}$ or an exponential disk (Arp 163).
\subsection{Concentration index}
The scale lengths derived from ellipse fitting depend on the structure
namely, the ellipticity and the position angle of the fitted isophotes.
In order to
compare the scale lengths etc. from morphologically dissimilar galaxies,
it is more accurate to use an index that does not change with the morphology.
To have a model-independent handle on the concentration of the light in the
galaxy, we derive the ratio of the radius that encloses 75$\%$ of the total
light to the radius that encloses 25$\%$ of the total light. The total
light was the light enclosed within an elliptical isophote at the
20 $mag\ arcsec^{-2}$ level and was taken from the 2MASS database.
This ratio is called the concentration index C$_{31}$ (de Vaucouleurs 1977)
 and is presented
in Table~\ref{props}. Plotting the frequency distribution of C$_{31}$
in the two classes (Fig. 6), we find that Class I objects show a peak at a value
between 4 and 5 while Class II objects show a double peaked distribution with a gap
in between. The gap corresponds to the region occupied by Class I objects.
We find that the highest C$_{31}$ objects are made up largely of galaxies
with a nuclear AGN/LINER component (except for Arp 224 and AM1315-263), as shown by the shading.
The left peak corresponding to a lower C$_{31}$ is due to non-active
objects. Thus, active galaxies having an outer exponential profile are found to
be more centrally concentrated as compared to the $r^{1/4}$ (completely relaxed)
systems which in turn are more centrally concentrated than non-active
galaxies having an outer exponential profile. Apparently, the AGN in the central region of the Class II galaxies contributes to a great extent to the
total light in the centre, as a result of which they appear more centrally
concentrated compared to the non-active Class II galaxies.

The fitted values of the bulge (R$_e$)and disk (R$_d$)
scale lengths and the surface brightnesses ($\mu_e$ and $\mu_0$ for Class I and Class II respectively) for the sample galaxies
obtained in Section 3.1 are given in Table~\ref{props}.
We used the velocity measurements from SIMBAD to derive the distances to the
galaxies and to compute the linear resolutions. A value of
H$_0$ = 75 $km\ s^{-1}\ Mpc^{-1}$ is used throughout.
 The derived quantities are listed in Table~\ref{props}.

\section{Results}
\subsection{Dynamical properties}
Despite their similar optical appearance and hence a similar stage of
dynamical evolution of tidal features, the galaxies seem to fall into two
distinct groups with different mass profiles. Thus, the galaxies in the
two classes have undergone a very different evolutionary pattern.
The natural physical explanation of the Class I galaxies is that each is
a largely relaxed merger of equal mass giant galaxies as studied
in typical numerical
simulations (e.g. Barnes \& Hernquist 1991). The origin of the Class II
galaxies is a puzzle. It may be that the galaxies in Class II
undergo a merger with a satellite or a minor merger with a mass ratio $\ge$ 0.1
and hence the exponential
disk in the larger galaxy is not significantly disturbed.
This conjecture for Class II is supported by the observations of
NGC 4424 by Kenney et al.(1996) who report that morphological features generally
associated with a merger are seen in this case and the outer part of the galaxy has a disk-like exponential light profile.
They argue that the apparent survival of the dominant exponential disk in this
galaxy suggests that it has experienced an intermediate mass ratio (0.1-0.5)
merger. Further, Bekki (1998b) proposes unequal mass mergers with ratios (0.3-0.5) as the
mechanism for the creation of S0 galaxies with outer exponential disks (also see Barnes 1998).
Thus, we propose that K$_s$ band photometry, in particular the radial profile,
 can be used as
a diagnostic to differentiate between cases of equal-mass and unequal-mass
merger remnants and to place a limit on the possible mass ratios in an
unequal-mass merger.

Our scenario for the origin of Class II galaxies is supported by numerical
simulations of minor mergers by Walker, Mihos \& Hernquist (1996). They
show that the final stage of a minor merger still resembles a disk galaxy
but has a thicker and hotter disk. The remnants older than 6 $\times$ 10$^8$
years (see their Fig. 5) appear remarkably similar to the optical
appearance of the sample we have chosen in our study. However, these authors
do not give the resulting radial mass distribution. Our work shows that
the exponential mass distribution in the outer disks, as seen from the
large range in radius over which it fits (Section 3.2) and the fact that the error bars
are small (Fig. 3),
shows that the exponential nature is robust for the Class II galaxies.

We note that the theoretical work so far has mostly concentrated
on studying a merger of equal mass galaxies (see Section 1), because the
focus has been mainly to prove that a merger of spirals {\it can} give
rise to an elliptical galaxy, though the unequal-mass encounters are
also now being studied (e.g. Barnes 1998).
The dynamical
evolution of merging galaxies resulting in an exponential outer profile
as observed in Class II is an open problem, and needs to be further studied by
future N-body work. It is puzzling that the second oldest remnant in the 
Toomre sequence, namely Arp 224, should show an exponential disk behaviour
in the outer region.
It is difficult to explain how a merging pair can have a disturbed 3-D light
distribution and yet have a smooth, exponential disk mass distribution, as in a
typical spiral galaxy.

Recently,
Naab \& Burkert (2001) have performed N-body simulations of merging disk
galaxies with mass ratios of 1:1,2:1,3:1 and 4:1. They predict the kinematic
and the photometric properties of the expected remnants. They study the
distribution of the relative position angle ($\Delta\phi$) between the isophote
at 0.5 $r_{eff}$and 1.5 $r_{eff}$ versus the effective ellipticity $\epsilon_{eff}$
(ellipticity at 1.5 $r_{eff}$),where $r_{eff}$ is the half-light radius, for random projections of each merger remnant.
Since our sample may cover different mass ratios viewed at different
inclination angles, it is meaningful to compare the observed values of
($\Delta\phi$) and $\epsilon_{eff}$ with their results to see if there is
an overlap. We use the half-light radii derived from the elliptical aperture photometry of
our sample and plot the behaviour of $\Delta\phi$ versus $\epsilon_{eff}$
for our sample of galaxies in Fig. 7. {\it We find that our observed values
lie in the same region of the plot as predicted theoretically by them.}
However, since most of the region is common to all the four mass ratios,
no clear comment can be made about the mass ratios of the progenitors.
Nevertheless, what is interesting is that a few of our objects clearly lie in the region corresponding to unequal mass mergers.
\subsection{Other properties}
Interestingly, the various classes in our study
do not show any clear demarcation in properties like
$F(100 \mu) / F(60 \mu)$ or the infrared luminosity. The ratio of
the infrared luminosity to the molecular hydrogen mass (taken from
Georgakakis et al. 2000)
does not also show any clear demarcation between the classes.
Also, there is no correlation between the class to which the galaxy belongs
and the blue luminosities or the K$_s$ band magnitudes.
Therefore, it is the ratio of the galaxy masses
that seems to be the main parameter that decides the class to
which a merger remnant belongs.

Some hints about the physical differences between these two classes could
be provided by the HI distribution in these systems since it traces
the dynamical consequences
of encounters at large radii.
Hibbard (2001) has compiled HI maps for peculiar
galaxies and galaxies showing peculiar HI distributions and divided them into
various groups. An inspection of these maps shows that seven of the objects
from our sample are included.
Interestingly, the three objects from Class I namely
Arp 193, Arp 226 and Arp 243 all fall in the same group of
interacting doubles made up of two HI systems with two HI tails.
Four objects in Class II fall in different groups.
Arp 186 is classified as interacting double made up of two HI systems with
one HI tail; Arp 214 has a two-sided HI warp, Arp 224 as interacting doubles
made up of two systems with only one containing HI and Arp 230 is a
peculiar early type galaxy with HI within the optical body.
This evidence points
to a homogeneous set of initial conditions as indicated by similar
HI properties for Class I. In contrast, Class II objects show a
diversity of HI behaviour, which points to varied initial
conditions for the galaxy encounter or with varied parameters for the
progenitor galaxies.

\subsection{Comparison with earlier work}
SB91 carried out a study similar to our work, namely a near infrared
study of advanced disk-disk mergers of galaxies chosen from the Arp sample. 
Six galaxies from their sample overlap with our sample. However, they
computed and fit the surface brightness profiles only along the major axis and
the minor axis and studied their deviations from the r$^{1/4}$ law
along these
two axes. Our study consists of fitting ellipses to the galaxy image and
deriving an averaged profile.
Thus, our study gives a better sampling of the galaxy brightness distribution.
We find that out of the six galaxies in common, five can be fit by an r$^{1/4}$ profile
while one galaxy, namely Arp 224, is fit better by an outer exponential.

Reshetnikov et al.(1993) studied a sample of 73 galaxies in
close interacting systems selected from the Arp
catalogue in the R band. Their sample concentrated on close double or triple
interacting systems and was selected based on the presence of obvious
morphological signs of interactions, such as bridges, tails, distortions,
envelopes, etc. They found that 10 galaxies show an r$^{1/4}$ profile, while 28 have
pure exponential brightness distributions and 30 galaxies show both components.
They report that bulges of interacting elliptical and S0 galaxies are more
compact compared to normal galaxies, while bulges of interacting spirals are
similar to those of non-interacting galaxies. Our study concentrates on advanced
merger candidates where the
systems have merged to a single nucleus stage. Hence, we are sampling
a more advanced stage in the merger process compared
to the Reshetnikov sample.

Zheng et al. (1999) studied a sample of 13 single-nucleus
ultraluminous galaxies using
HST images. They find a nearly equal number of galaxies that can be fitted by an
$r^{1/4}$, inner $r^{1/4}$ plus an outer extension and profiles that
deviate from the $r^{1/4}$ law.
Scoville et al. (2000) studied a sample of 24 luminous infrared galaxies and found that 9 of them could be approximately fit by a $r^{1/4}$ law.
They also report that a majority of the systems (13 out of 24) could
be fit equally well by either an $r^{1/4}$ or an exponential law.
The result from our paper, namely that a merger need not always lead to a luminosity profile like that of elliptical
galaxies, is consistent with the results obtained by these authors, except for
a different sample.

We studied the distribution of the concentration index C$_{31}$ with
M$_{K_s}$ (the absolute K$_s$ magnitude) (Fig. 8) and found that
all the galaxies that can be fit by an r$^{1/4}$ law are brighter than
-21.0$^m$ and have a concentration index greater than 3.5.
We also find that there is a trend for lower luminosity galaxies to
have smaller values of C$_{31}$. A similar study by Boselli et al. (1997) of
Virgo cluster galaxies showed that C$_{31}$ $\leq$ 3 for all galaxies
fainter than 10.0$^m$ in K$^{\prime}$, independent of the morphological
type of the galaxy.They argue that this is consistent with an exponential disk.
Similarly, Gavazzi et al. (1996) found that luminous galaxies are more centrally concentrated.

\section{Conclusions}
A study of a sample of morphologically selected disturbed galaxies
with indications of mergers shows that
in spite of their similar optical appearances, the luminosity and hence
mass profiles of these
galaxies fall into two distinct classes. Thus, the dynamical evolution is distinctly different in the two classes.
Class I is made up of galaxies that are fit by an $r^{1/4}$ law and
hence are completely relaxed systems.
Class II objects show the presence of an exponential disk in their outer
regions. A few objects cannot be fit by either law. Thus, the dynamical
properties of these classes are distinct.
These classes do not show any differences in various properties, such as IR
luminosities, molecular gas content or the IR colours. We find that
our sample of morphologically disturbed galaxies may be merger products in spite of not being bright in the IR.
Class I objects are likely to have formed as a result of merging of equal
mass galaxies and
have relaxed completely.
Class II objects are likely to be the result of unequal mass mergers in which
the exponential disk of the primary progenitor has survived. The unrelaxed
objects
are likely to be recent mergers.

We have also studied the structural parameters such as B4 and
the location of the
centre of mass, that provides important information on mergers, in addition
to the mass distribution as indicated by the luminosity profiles.
It is seen that the concentration index C$_{31}$ for Class I objects shows
a maximum between 4 and 5. For Class II objects, there is a clear demarcation between the active and the non-active galaxies. The active galaxies show a
peak between 5 and 5.5 while the non-active galaxies show a peak between 3 and 3.5. Thus, it is seen that active galaxies
with an exponential disk are more centrally concentrated than
completely relaxed systems which in turn are more centrally concentrated
than non-active galaxies with an outer exponential disk.

In summary, 13 out of 27 disturbed galaxies with indications of mergers
are found to exhibit
exponential disk behaviour. Hence,
the existence of this class is a robust result and this class of galaxies
requires further dynamical study.

\begin{acknowledgements}
We are very thankful to the referee, R.A. Koopmann for detailed and
constructive comments which have greatly improved the paper.

This publication makes use of data products from the Two Micron All Sky Survey,
which is a joint project of the University of Massachusetts and the
Infrared Processing and Analysis Center/California Institute of Technology,
funded by the National Aeronautics and Space Administration and the National
Science Foundation.
This research has made use of the NASA/ IPAC Infrared Science Archive,
which is operated by the
Jet Propulsion Laboratory, California Institute of Technology, under
contract with the National
Aeronautics and Space Administration,
the NASA/IPAC Extragalactic Database (NED)
which is operated by the Jet Propulsion Laboratory, California Institute
of Technology, under contract with the National Aeronautics and Space
Administration,
the SIMBAD database, operated at CDS,
Strasbourg, France, the LEDA and the
HYPERCAT databases and NASA's Astrophysics
Data System Bibliographic Services. 
\end{acknowledgements}
\appendix
\section
{Notes on individual galaxies}
The structural parameters deduced for the galaxies in our sample
(Section 3.3) are listed below:

{\it Arp 156} : Low ellipticity, isophotal twisting

{\it Arp 165} : slightly disky, nearly constant P.A.

%{\it Arp 168} : Oscillating B4, nearly constant P.A.

{\it Arp 193} : Disky throughout, nearly constant P.A.

{\it Arp 221} : Inner disky, outer boxy between 10\arcsec\ and 20\arcsec\ .

{\it Arp 222} : Disky up to 30\arcsec\ (between 0.6 to 3.7 kpc), outer boxy

{\it Arp 225} : No clear trends, nearly constant P.A.

{\it Arp 226} : Very disturbed profile

{\it Arp 231} : Boxiness beyond 8\arcsec\ (between 2.9 and 5.8 kpc), P.A. keeps changing erratically

{\it Arp 243} : Inner disky, outer boxy (beyond 2.1 kpc). centre starts changing beyond 2.1 kpc.

{\it Arp 162} : Disky beyond 15\arcsec\ (1.3 kpc), nearly constant P.A.

{\it Arp 186} : Strong isophotal twisting, inner disky (up to 2.4 kpc), outer boxy

{\it Arp 187} : Slightly disky throughout, small changes in P.A

{\it Arp 212} : Disky, changing centre, isophotal twisting

{\it Arp 214} : Disky, erratically changing centre

{\it Arp 219} : Inner disky (between 2.6 and 4.7 kpc), boxy beyond

{\it Arp 224} : Isophotal twisting, erratic B4

{\it Arp 230} : Disky up to about 1.9 kpc, isophotal twisting

{\it AM0318-230} : Inner boxy (up to 4.8 kpc), outer disky (between 4.8 and 8.9 kpc), isophotal twisting

{\it AM0337-312} : Inner boxy (between 2.9 and 5.2 kpc), highly inclined, nearly constant PA.

{\it AM1324-431} : Inner boxy

{\it AM1315-263} : Isophotal twisting, boxy beyond 25\arcsec\ 

{\it AM2146-350} : Inner boxy, outer disky, isophotal twisting

{\it AM0501-632} : Disky, isophotal twisting. The only galaxy fit by an
exponential throughout.

{\it Arp 160} : Inner disky, outer boxy, continuously changing centre

{\it Arp 163} : Disky, continuously changing centre

{\it AM1300-233} : Inner disky, outer boxy, changing centre

{\it AM1306-282} : Disky up to 10\arcsec\ , changing centre

\newpage
\section{List of figures}
\begin{enumerate}
\item K$_s$ band contours of the sample galaxies. The contours are plotted at intervals of 0.25$^m$ with the faintest contour level being 19$^m$ for all the galaxies except for Arp 193, Arp 221, Arp 231 and Arp 243 for which the faintest contour is at 19.5$^m$ and at 18.5$^m$ for Arp 225. The X and Y axes are given in arcsecs.

\item Class I: Galaxies well fit by an $r^{1/4}$ law. The K$_s$ band magnitude in $mag\ arcsec^{-2}$ is plotted against the semi-major axis, $a$ \arcsec.

\item Class II : Galaxies well fit by an outer exponential disk. The K$_s$ band magnitude in $mag\ arcsec^{-2}$ is plotted against the semi-major axis, $a$ \arcsec.

\item Galaxies that could not be fit. The K$_s$ band magnitude in $mag\ arcsec^{-2}$ is plotted against the semi-major axis, $a$ \arcsec.

\item A4, B4, X0 and Y0 for the Class I galaxy Arp 193 and the unrelaxed galaxy Arp 163. The inner isophotes have a common centre for the relaxed system, Arp 193, whereas in contrast, the centre shows a wandering or a sloshing behaviour for the unrelaxed system Arp 163.

\item Distribution of the number of galaxies versus the concentration indices C$_{31}$ of the sample. The figure on the left denotes class I objects and the figure on the right denotes class II. The shaded region comprises of objects which are AGN/LINERs.

\item Distribution of the relative position angle $\Delta\phi$ between 0.5 $r_{eff}$ and 1.5 $r_{eff}$ with the ellipticity $\epsilon_{eff}$ at 1.5 $r_{eff}$. Class I objects are denoted by circles, Class II by crosses and the unrelaxed objects by squares. Our observed values lie in the same region of the plot as predicted theoretically for mergers by Naab \& Burkert (2001).

\item The concentration index as a function of the absolute K$_s$ magnitude (M$_{K_s}$). The open circles denote Class I, the crosses denote Class II and the squares denote the unrelaxed objects.

\end{enumerate}

\begin{thebibliography}{}
%\bibitem[1966]{arp} Arp, H. C. 1966, ApJS, 14, 1
\bibitem[1966]{arp} Arp, H. C. 1966, {\it Atlas of Peculiar Galaxies}, Pasadena:
California Institute of Technology)
\bibitem[1987]{am} Arp, H. C., \& Madore, B. F. 1987, {\it A Catalog of Southern Peculiar Galaxies and Associations}, Cambridge Univ. Press, Cambridge
%\bibitem[] {am} Arp, H. C., Madore, B. F. 1975, Obs, 95, 212
\bibitem[1998]{baggett} Baggett, W.E., Baggett, S.M., \& Anderson, K.S.J. 1998, AJ, 116, 1626
\bibitem[1988] {barnes88} Barnes, J.E. 1988, ApJ, 331, 699
\bibitem[1998] {barnes98} Barnes, J.E. 1998, in Interactions and Induced Star Formation: Saas-Fee Advanced Course 26, eds. D. Friedli et al. (Berlin: Springer Verlag), pg. 275
\bibitem[1991] {barnes} Barnes, J.E., \& Hernquist, L.E. 1991, ApJ, 370, L65
\bibitem[1998a]{bekki1} Bekki, K. 1998a, ApJ, 496, 713
\bibitem[1998b]{bekki} Bekki, K. 1998b, ApJ, 502, L133
%\bibitem[1993]{bern} Bernloehr, K. 1993, A\&A, 268, 25
\bibitem[1998]{bm} Binney, J., \& Merrifield, M., 1998, Galactic Astronomy, Princeton: Princeton Univ. Press
\bibitem[1997]{boselli} Boselli, A., Tuffs, R.J., Gavazzi, G., Hippelein, H., \& Pierini, D. 1997, A\&AS, 121, 507
\bibitem[1977]{deV} de Vaucouleurs, G. 1977, in Evolution of Galaxies and Stellar Populations, ed. B.M. Tinsley, \& R.B. Larson (New Haven:Yale University Obs.), p. 43
\bibitem[1996]{gavazzi} Gavazzi, G., Pierini, D., \& Boselli, A. 1996, A\&A, 312, 397 
\bibitem[2000]{geor}  Georgakakis, A., Forbes, D.A.,  \& Norris, R.P. 2000, MNRAS, 318, 124.
\bibitem[1992]{hern}   Hernquist, L.E. 1992, ApJ, 400, 460
\bibitem[1992]{hernsper} Hernquist, L., \& Spergel, D. N. 1992, ApJ, 399, L117
\bibitem[2001]{hibbard} Hibbard, J.E., van Gorkom, J.H., Rupen, M.P., \& Schiminovich, D. 2001, astro-ph/0110667
\bibitem[1999]{james} James, P., Bate, C., Wells, M., Wright, G., \& Doyon, R. 1999, MNRAS, 309, 585
\bibitem[1987] {jedr} Jedrejewski, R.I. 1987, MNRAS, 226, 747
\bibitem[1985]{joseph} Joseph, R.D., \& Wright, G.S. 1985, MNRAS, 214, 87
\bibitem[1996]{kenney} Kenney, J.D.P., Koopmann, R.A., Rubin, V.C., \& Young, J.S. 1996, AJ, 111, 152
\bibitem[1998]{lu} Lu, N.Y. 1998, ApJ, 506, 673
\bibitem[1967]{lynden} Lynden-Bell, D. 1967, MNRAS, 136, 101
\bibitem[2001]{naab} Naab, T., \& Burkert, A. 2001, astro-ph/0110179
\bibitem[1989]{nieto} Nieto, J.-L., \& Bender, R. 1989, A\&A, 215, 266
\bibitem[1993]{resh} Reshetnikov, V.P., Hagen-Thorn, V.A. \& Yakovleva, V.A. 1993, A\&AS, 99, 257
%\bibitem[1988]{sanders}  Sanders, D.B. 1988, ApJ, 325, 74
\bibitem[1996]{sm} Sanders, D.B., \& Mirabel, I.F. 1996, ARA\&A, 34, 749
\bibitem[1996]{schweizer} Schweizer, F. 1996, AJ, 111, 109
\bibitem[1982]{sch}  Schweizer, F. 1982, ApJ, 252, 455
%\bibitem[1992]{sch2} Schweizer, F., \& Seitzer, P. 1992, AJ, 104, 1039
\bibitem[2000]{scov} Scoville, N.Z., Evans, A.S., Thompson, R., Rieke, M., Hines, D.C., Low, F.J., Dinshaw, N., Surace, J.A., \& Armus, L. 2000, AJ, 119, 991
%\bibitem[1998]{shi} Shier, L.M., \& Fischer, J. 1998, ApJ, 497, 163
\bibitem[1991]{stan} Stanford, S.A., \& Bushouse, H.A., 1991, ApJ, 371, 92
\bibitem[1977]{toom} Toomre, A. 1977, in The Evolution of Galaxies and Stellar Populations, ed. B.M. Tinsley, \& R.B. Larson (New Haven:Yale University Obs.), 401.
\bibitem[1982] {van} van Albada, T.S. 1982, MNRAS, 201 939
\bibitem[1996] {walker} Walker, I. R., Mihos, J.C., \& Hernquist, L. 1996, ApJ, 460, 121
\bibitem[1990]{wright} Wright, G.S., James, P.A., Joseph, R.D., \& McLean, I.S., 1990, Nature, 344, 417
\bibitem[1999]{zheng} Zheng, Z., Wu, H., Mao, S., Xia, X.-Y., Deng, Z.-G., \& Zou, Z.-L. 1999, A\&A, 349, 735
\end{thebibliography}
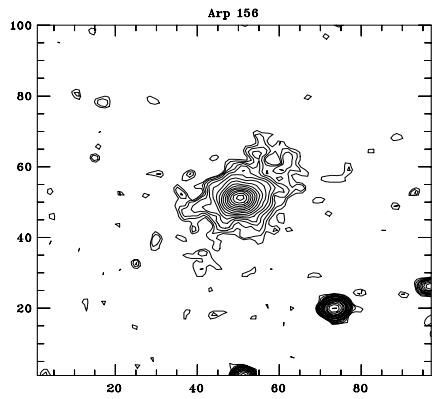
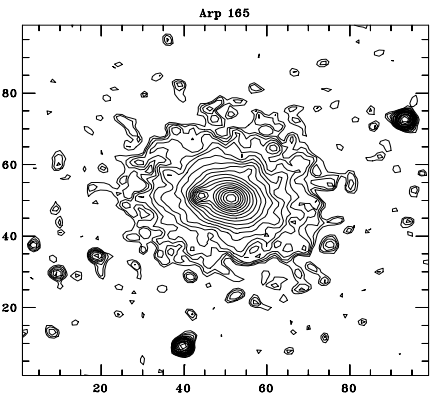
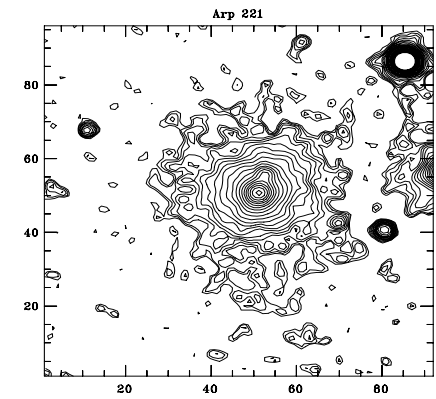
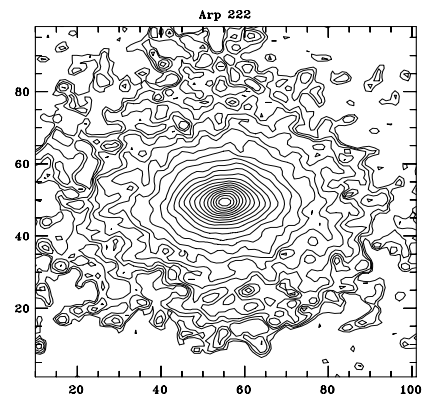
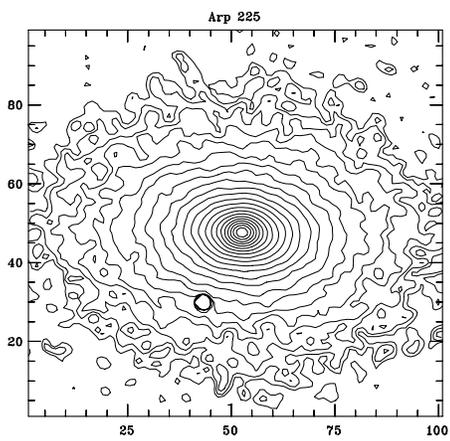
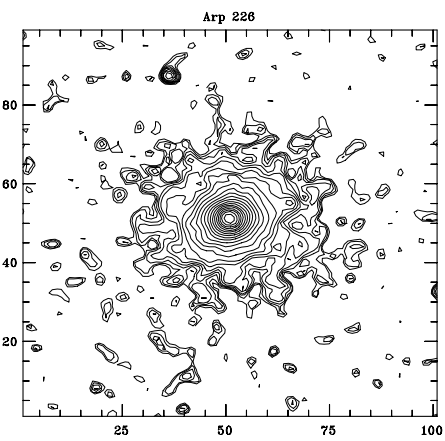
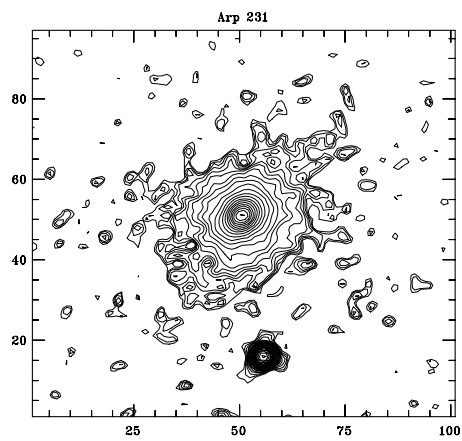
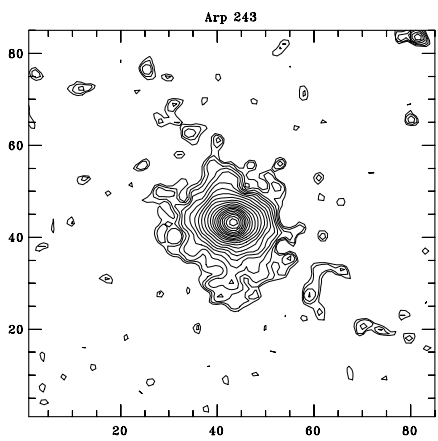
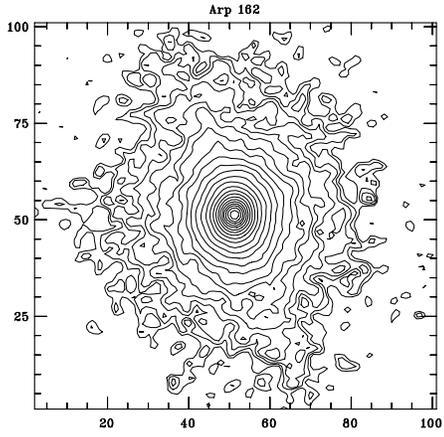
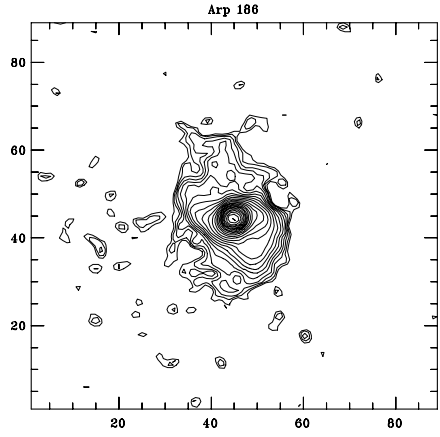
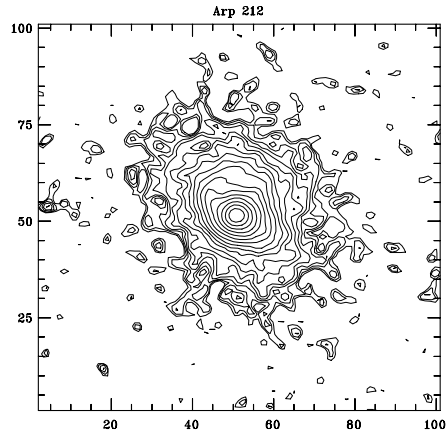
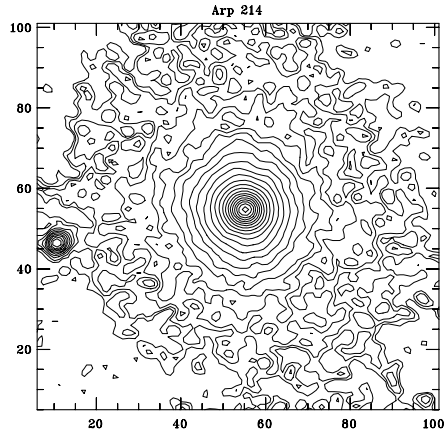
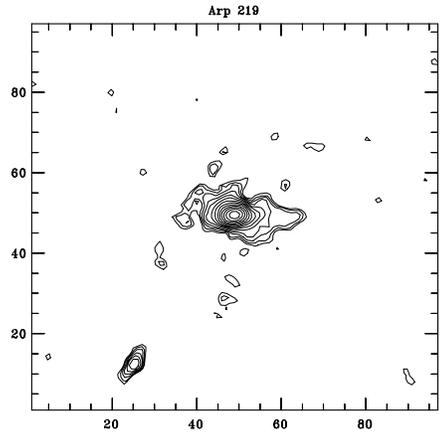
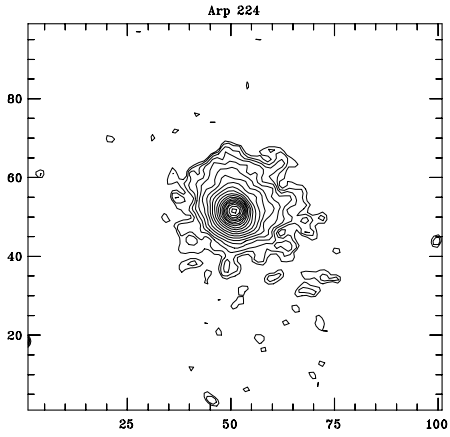
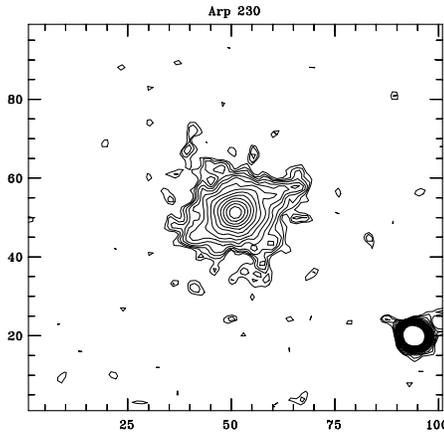
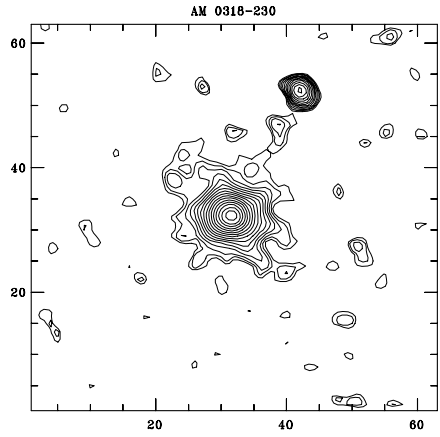
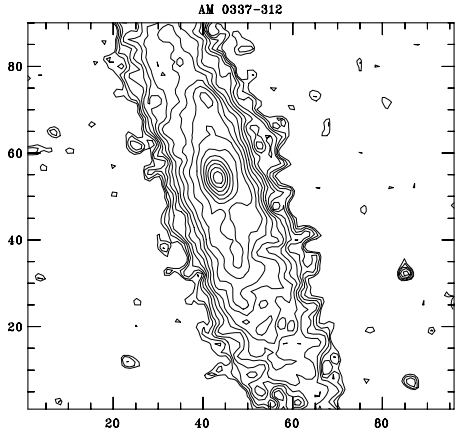
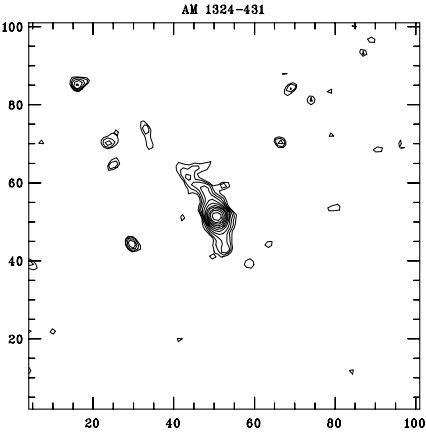
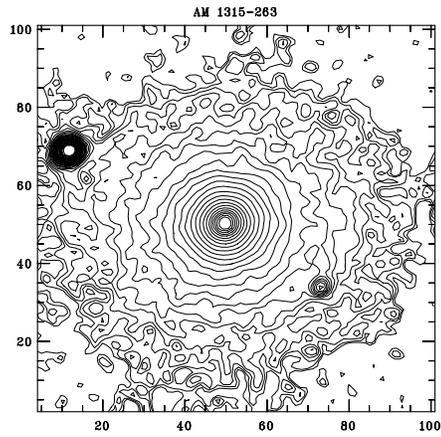
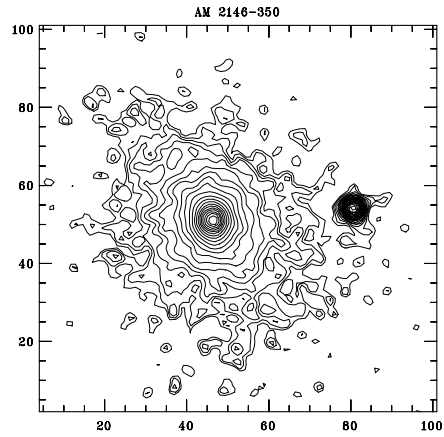
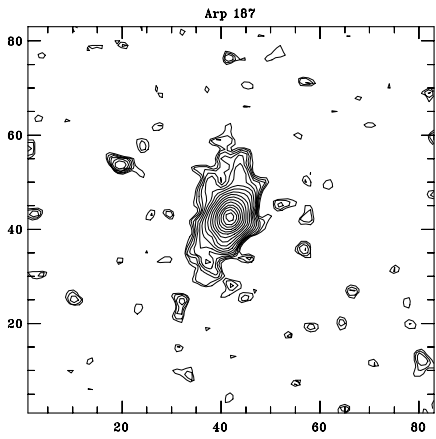
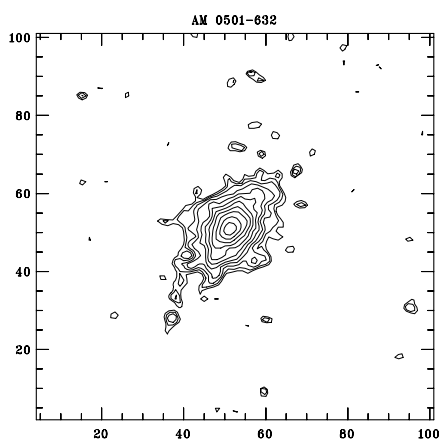
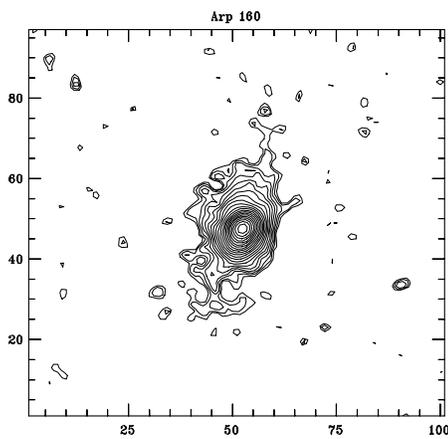
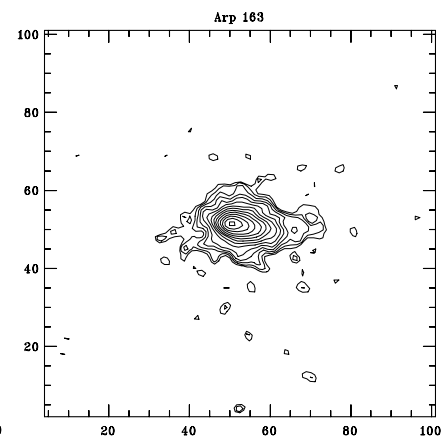
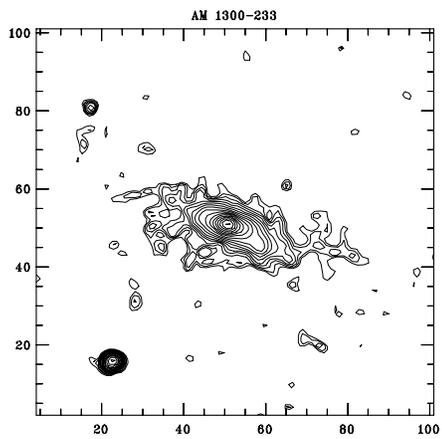
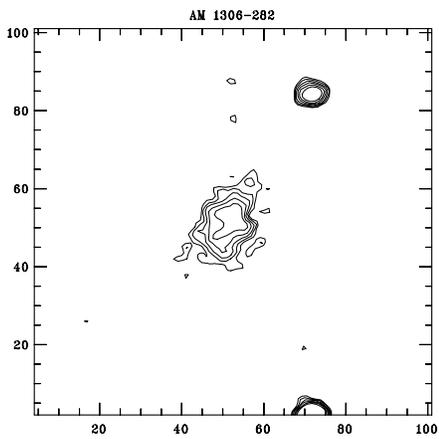
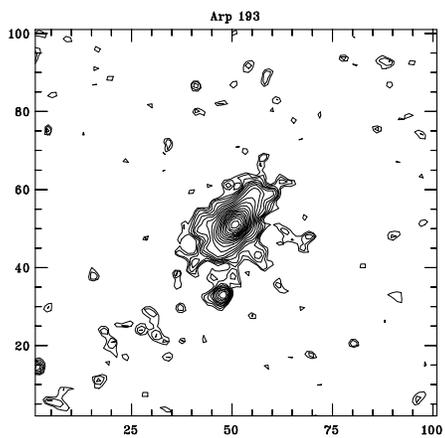
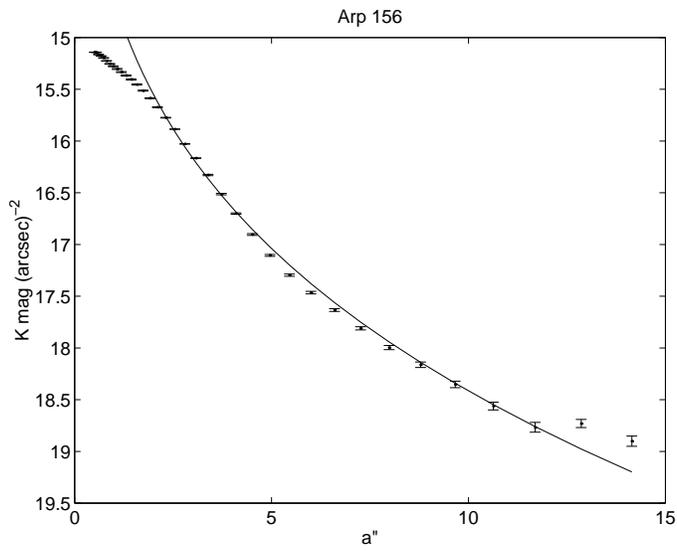
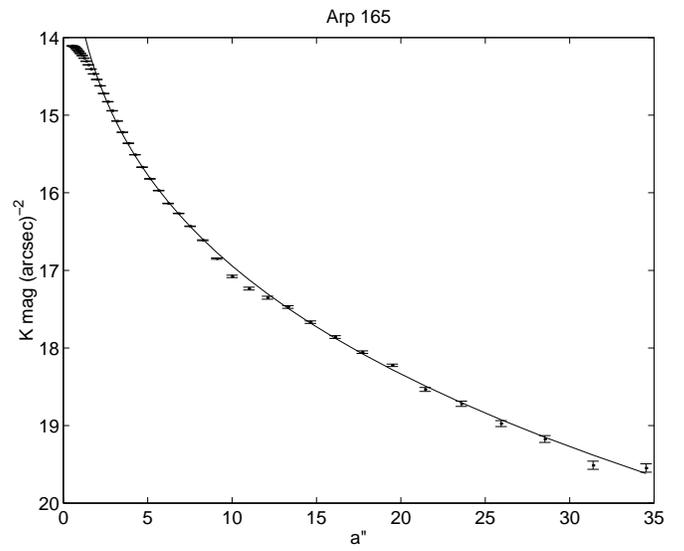
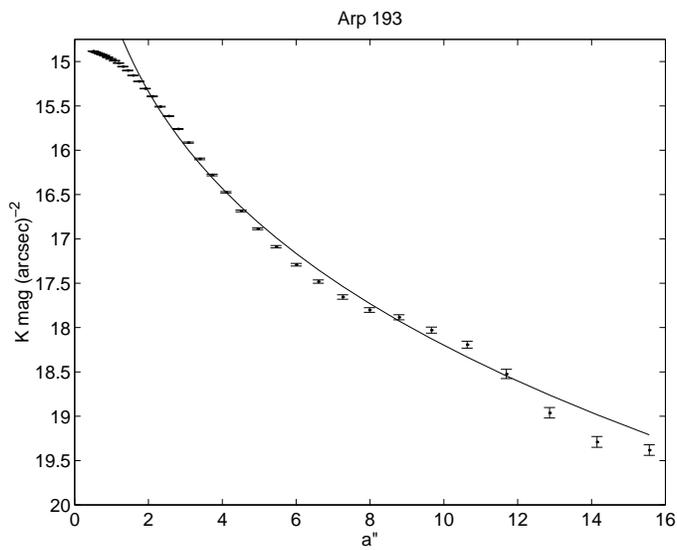
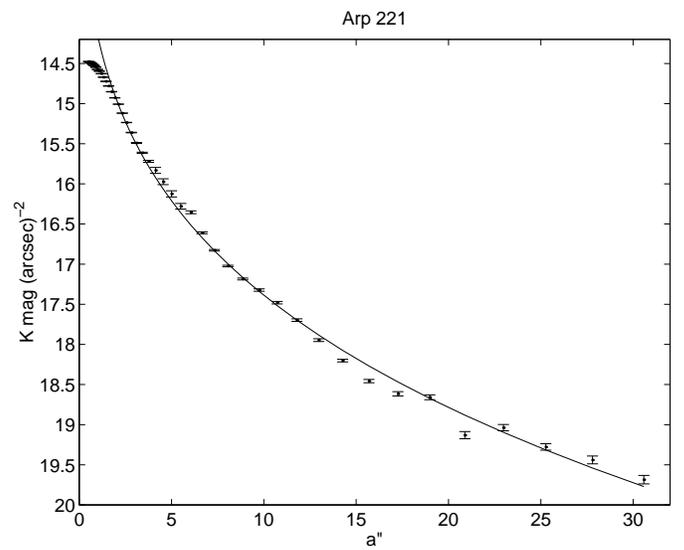
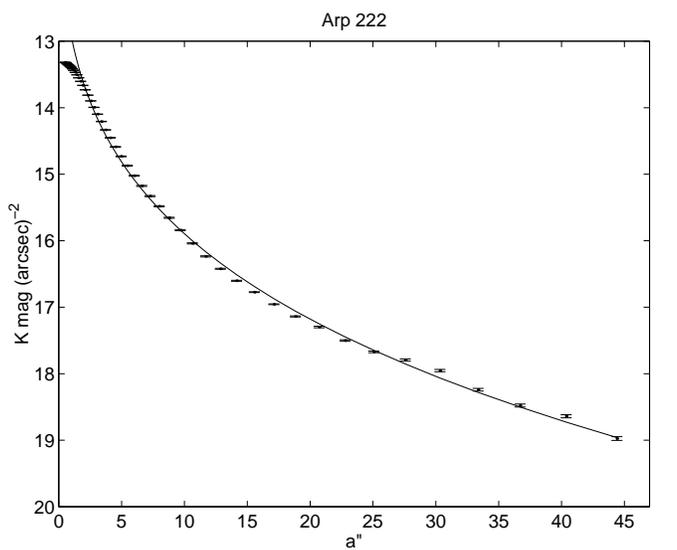
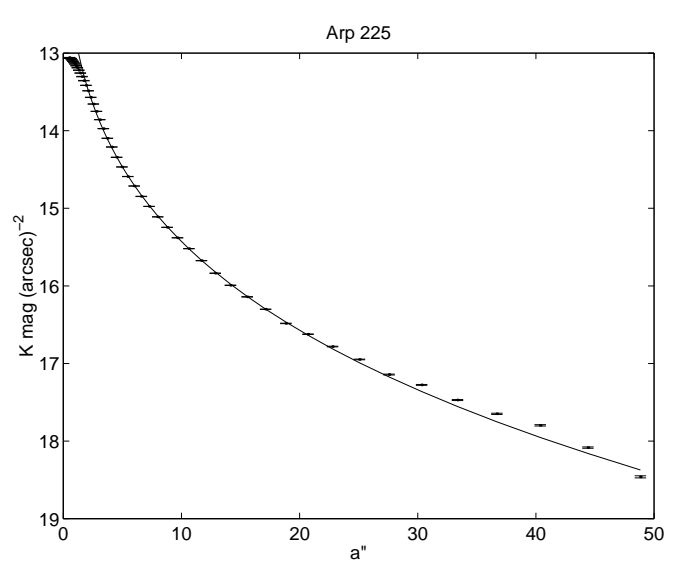
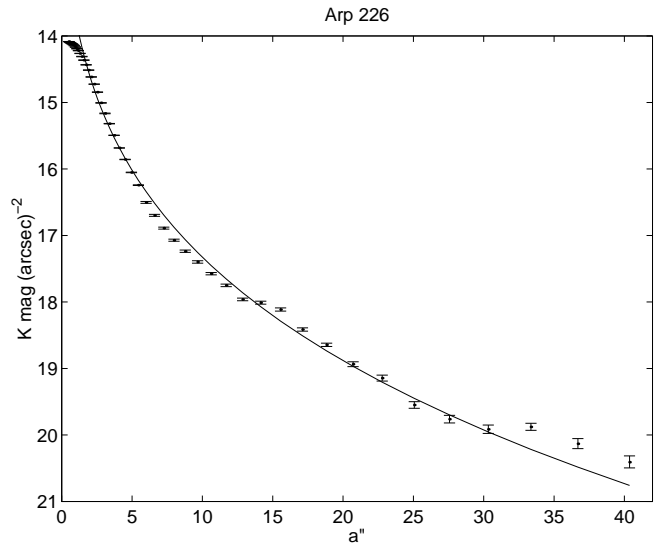
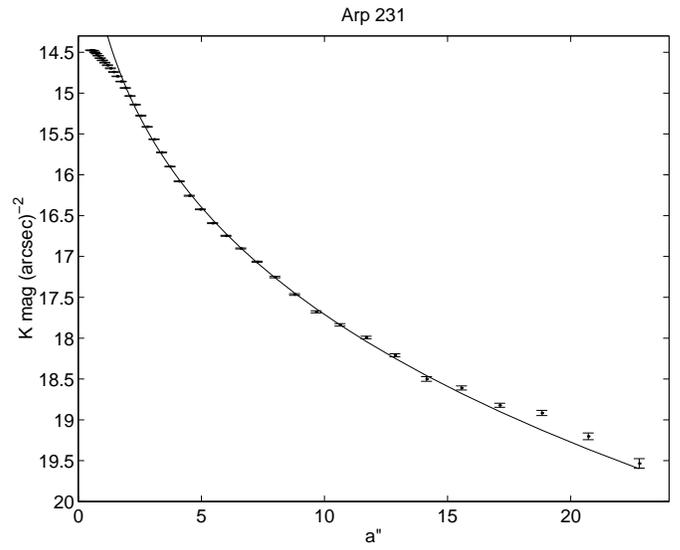
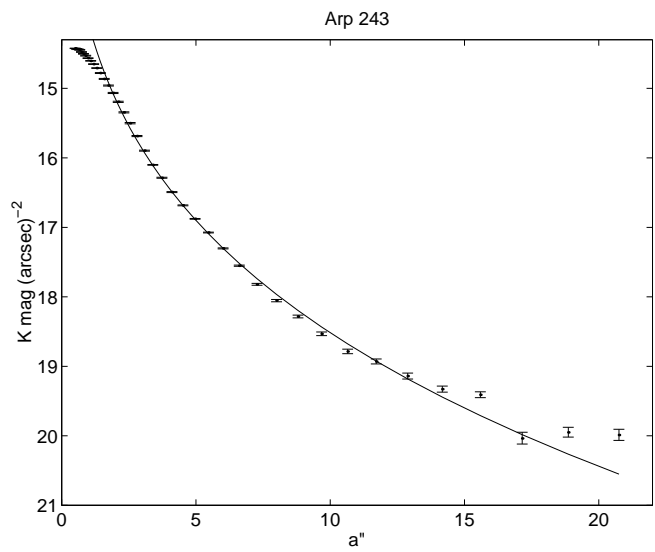
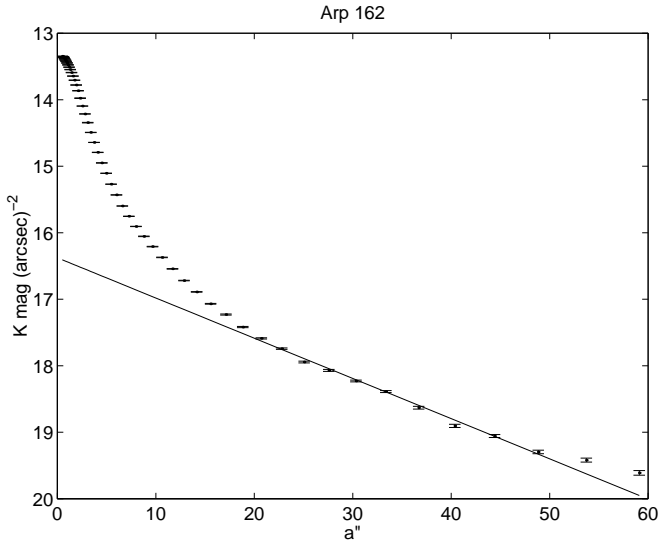
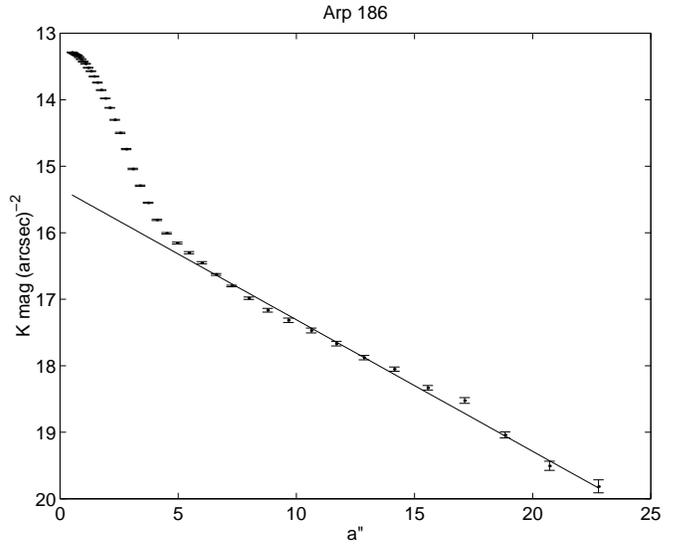
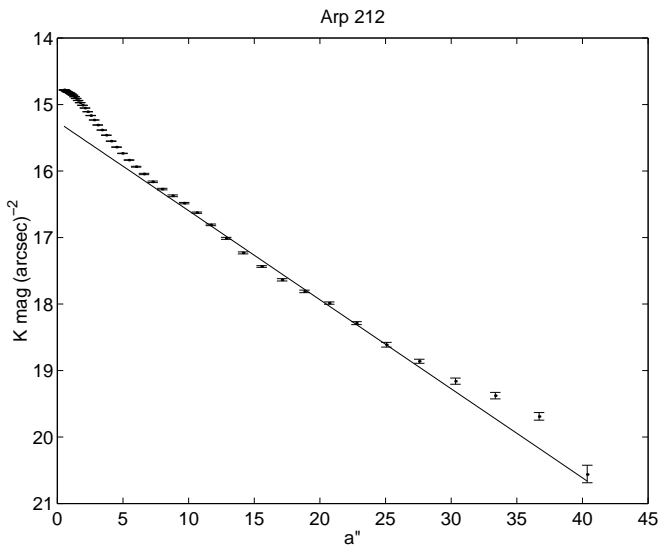
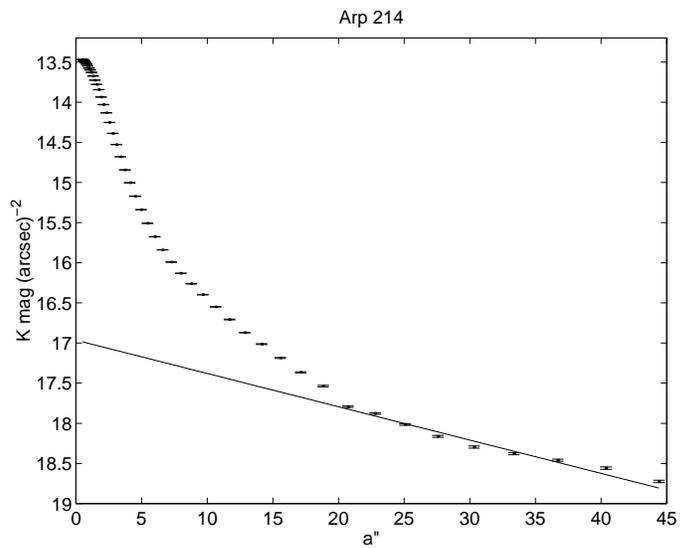
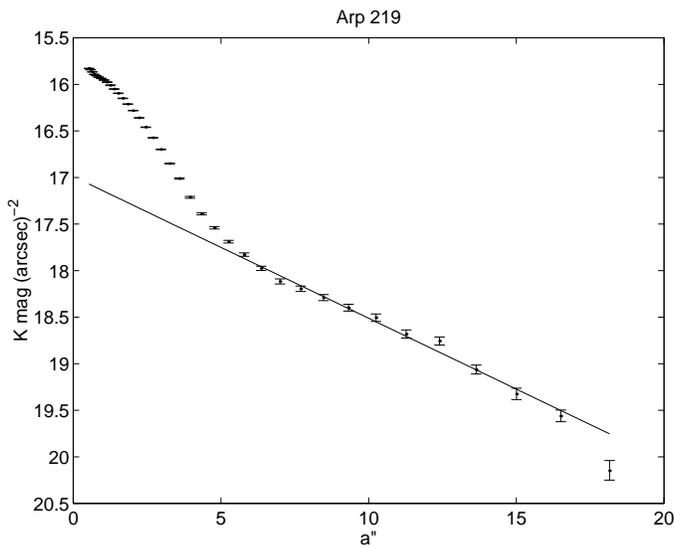
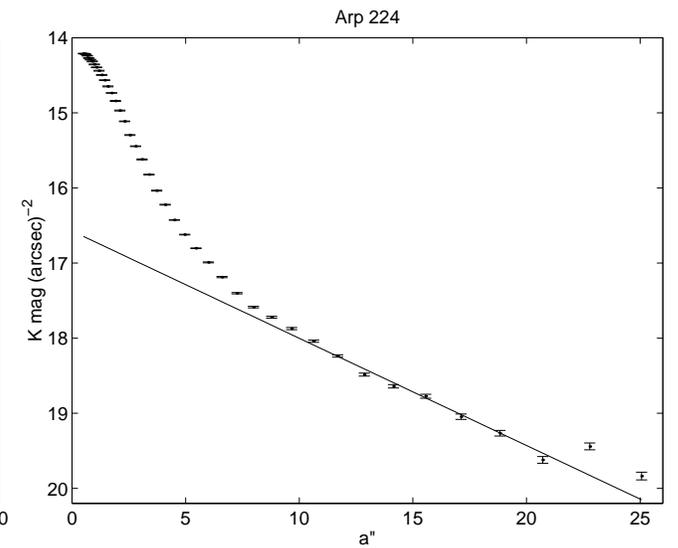
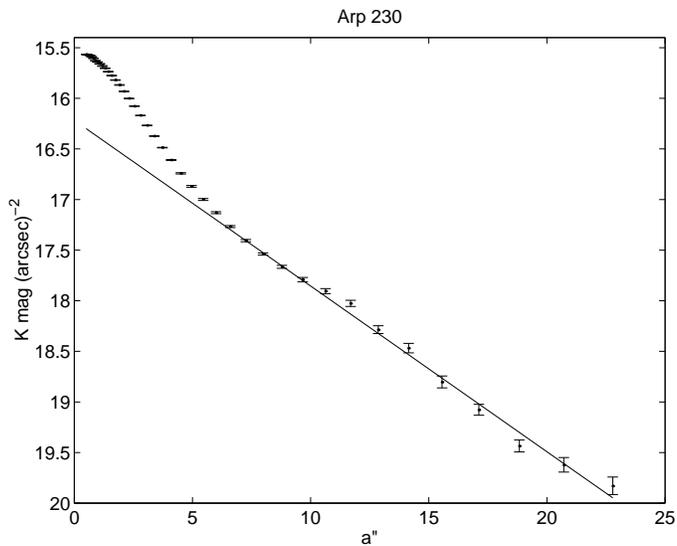
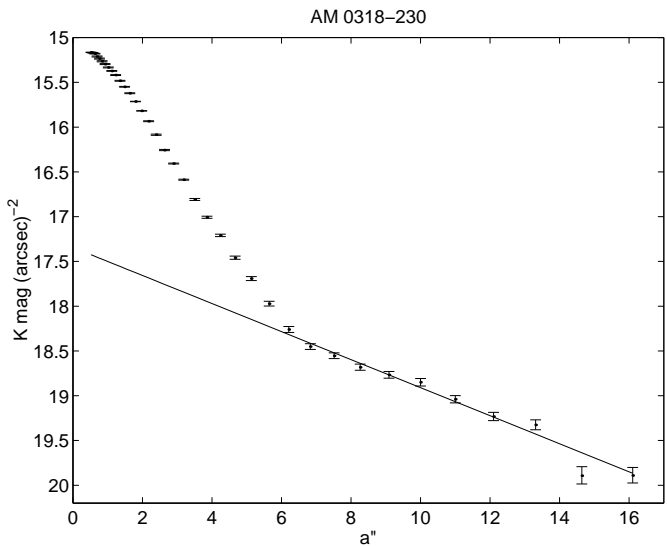
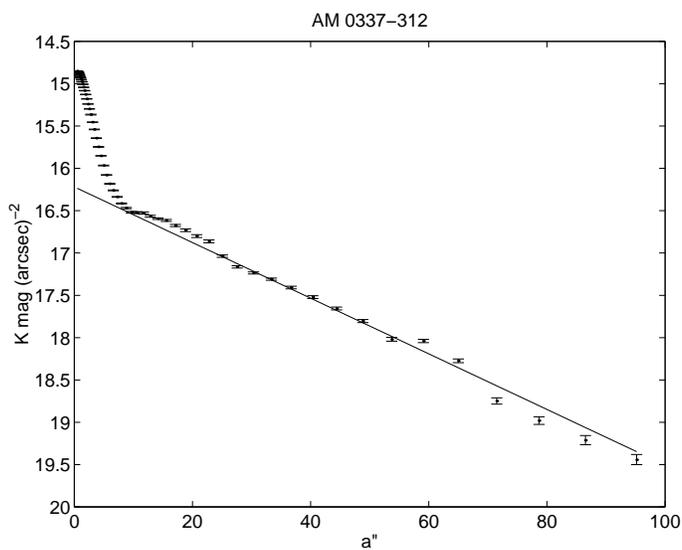
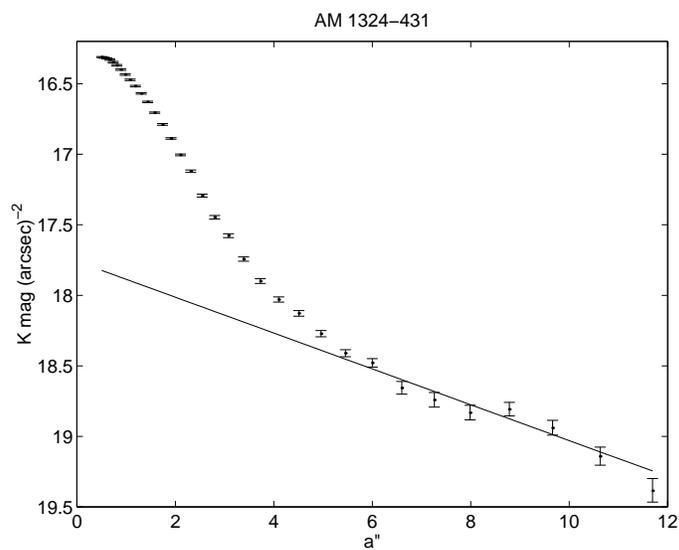
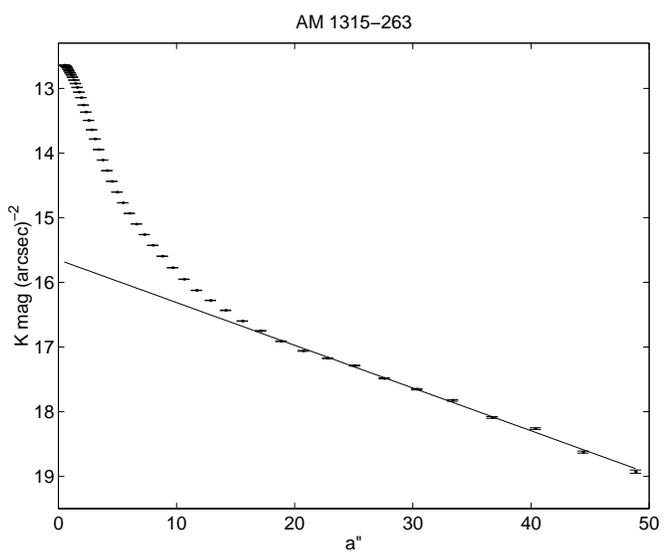
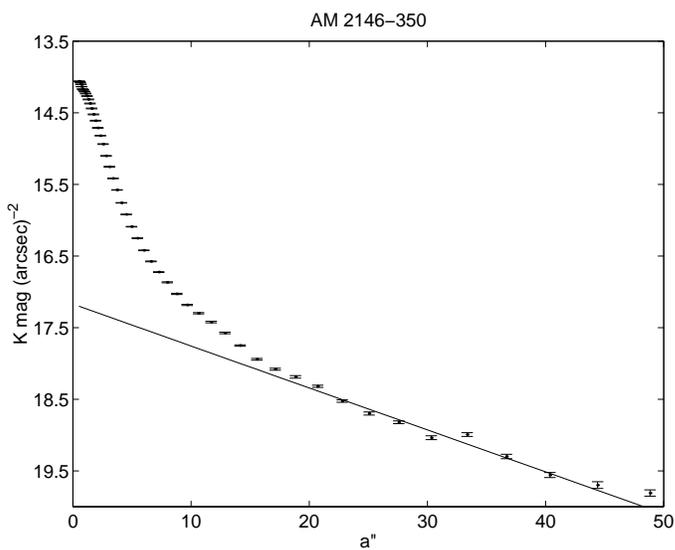
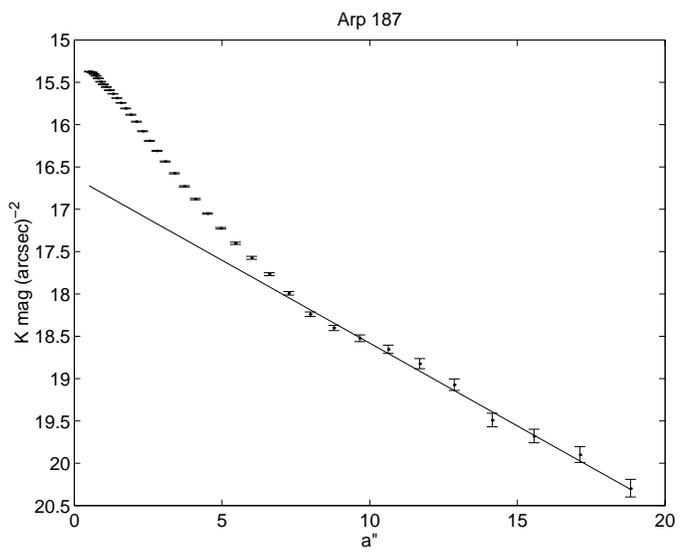
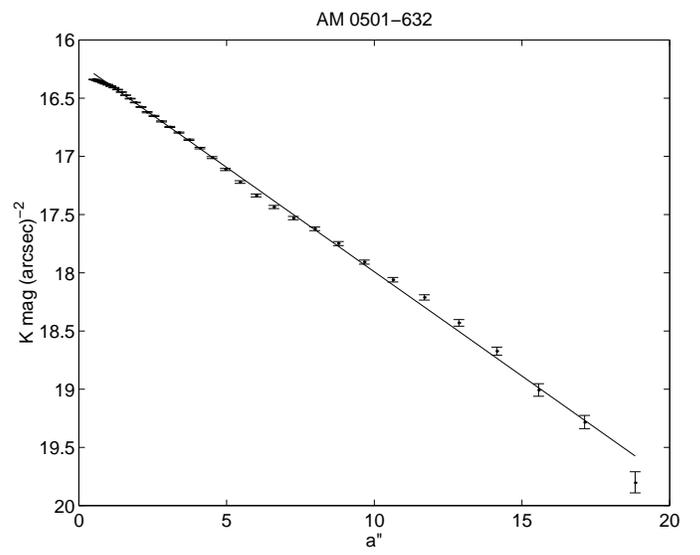
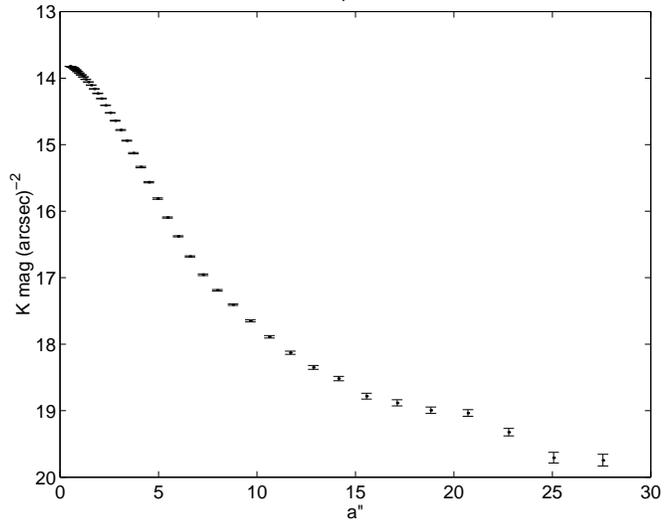
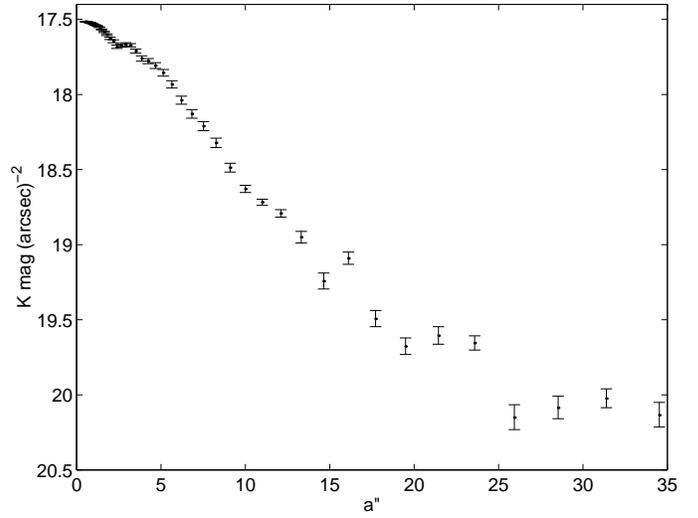
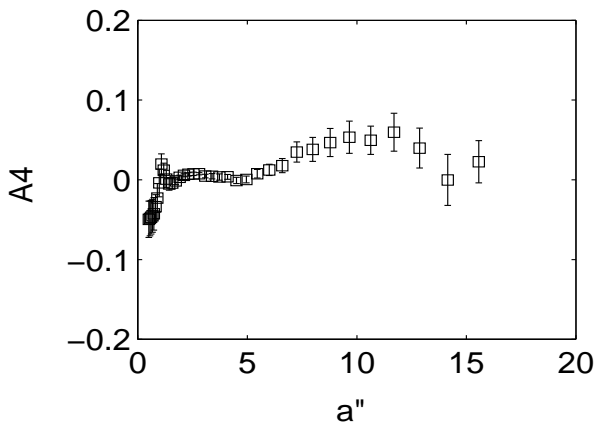
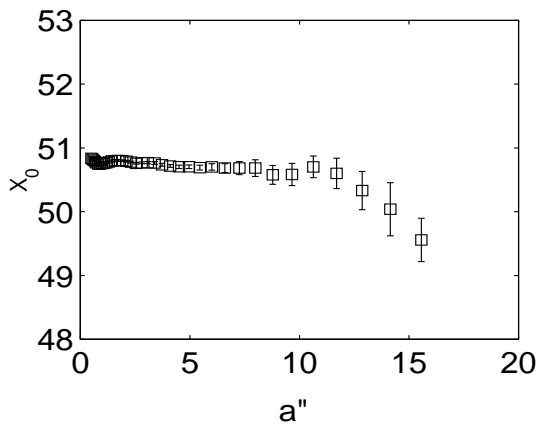
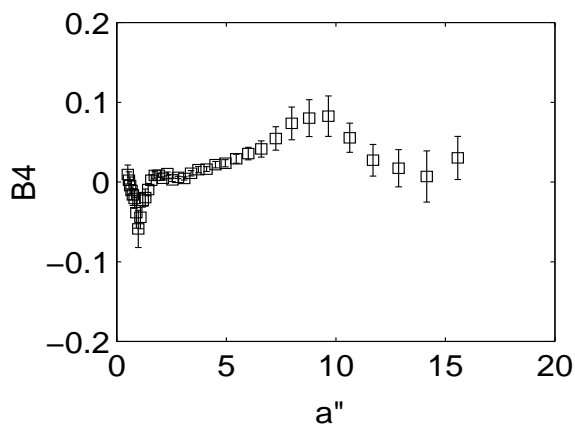
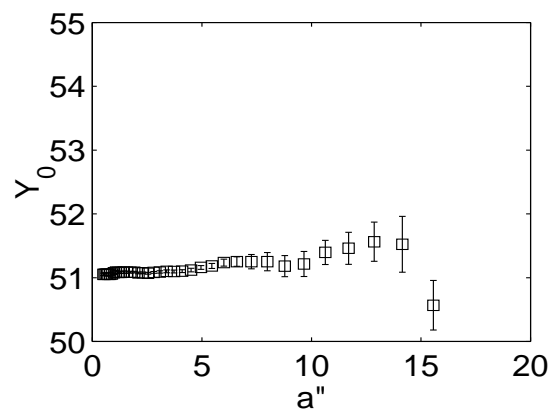
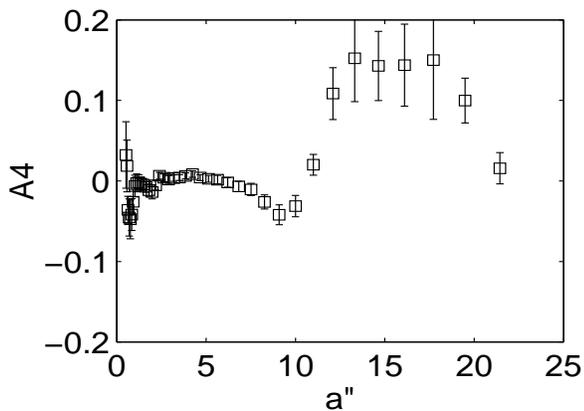
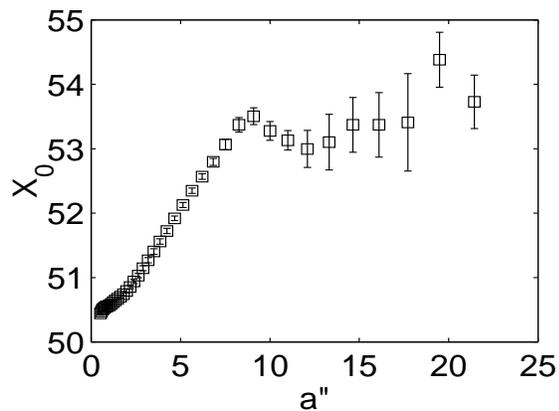
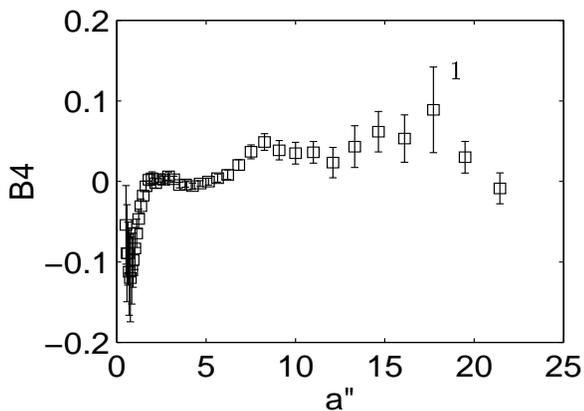
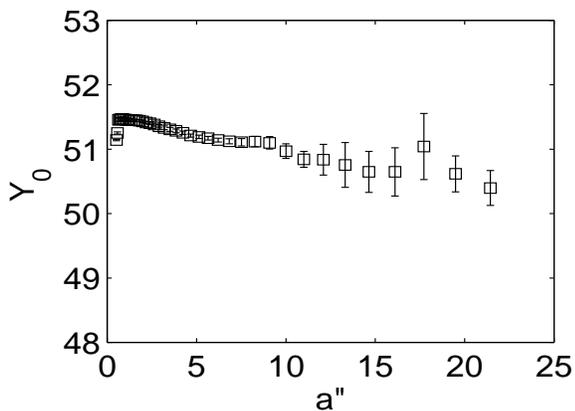
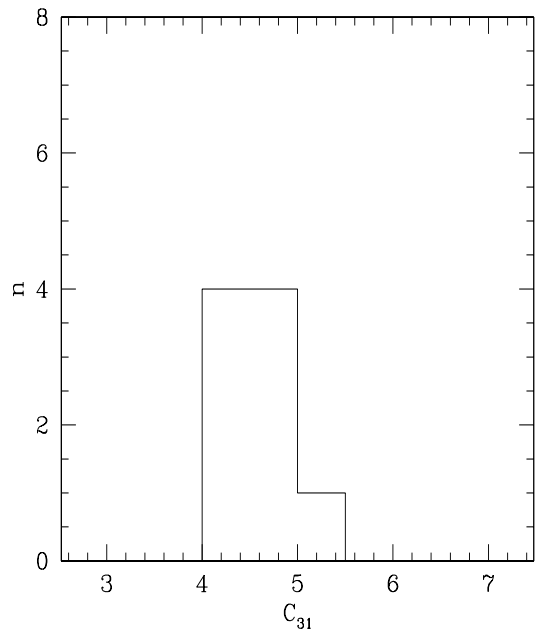
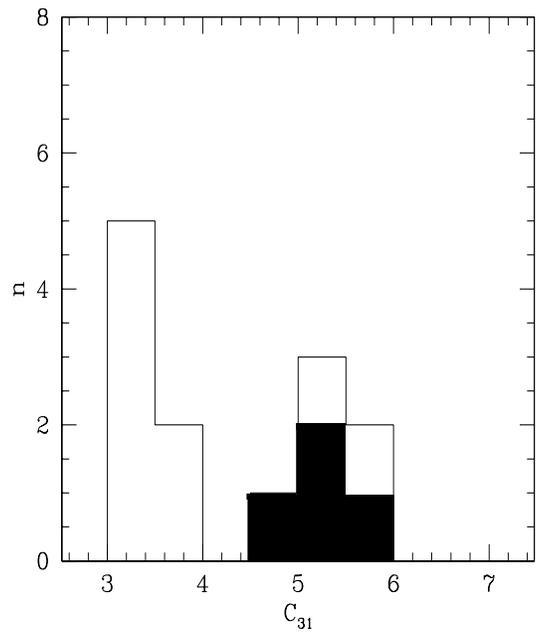
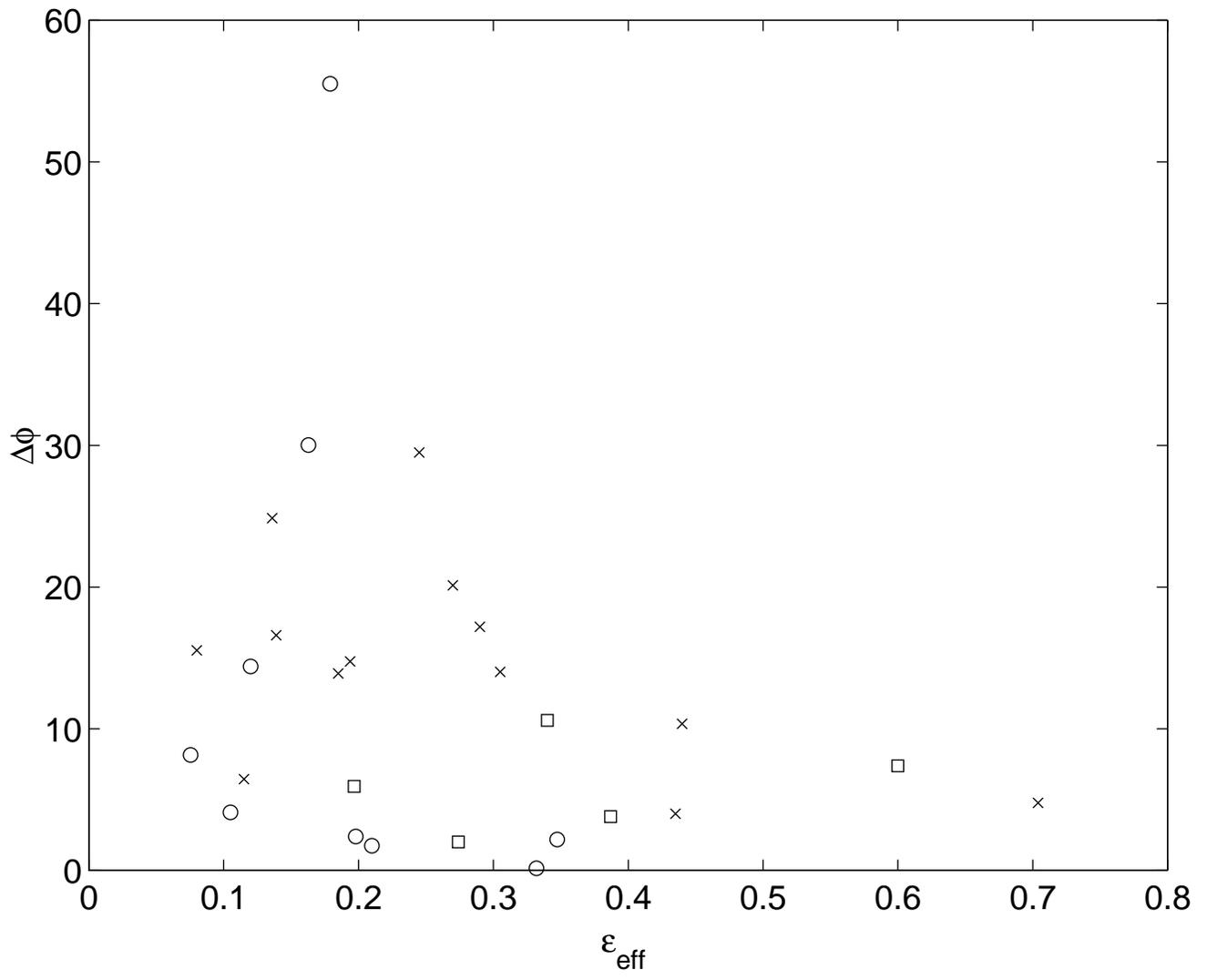
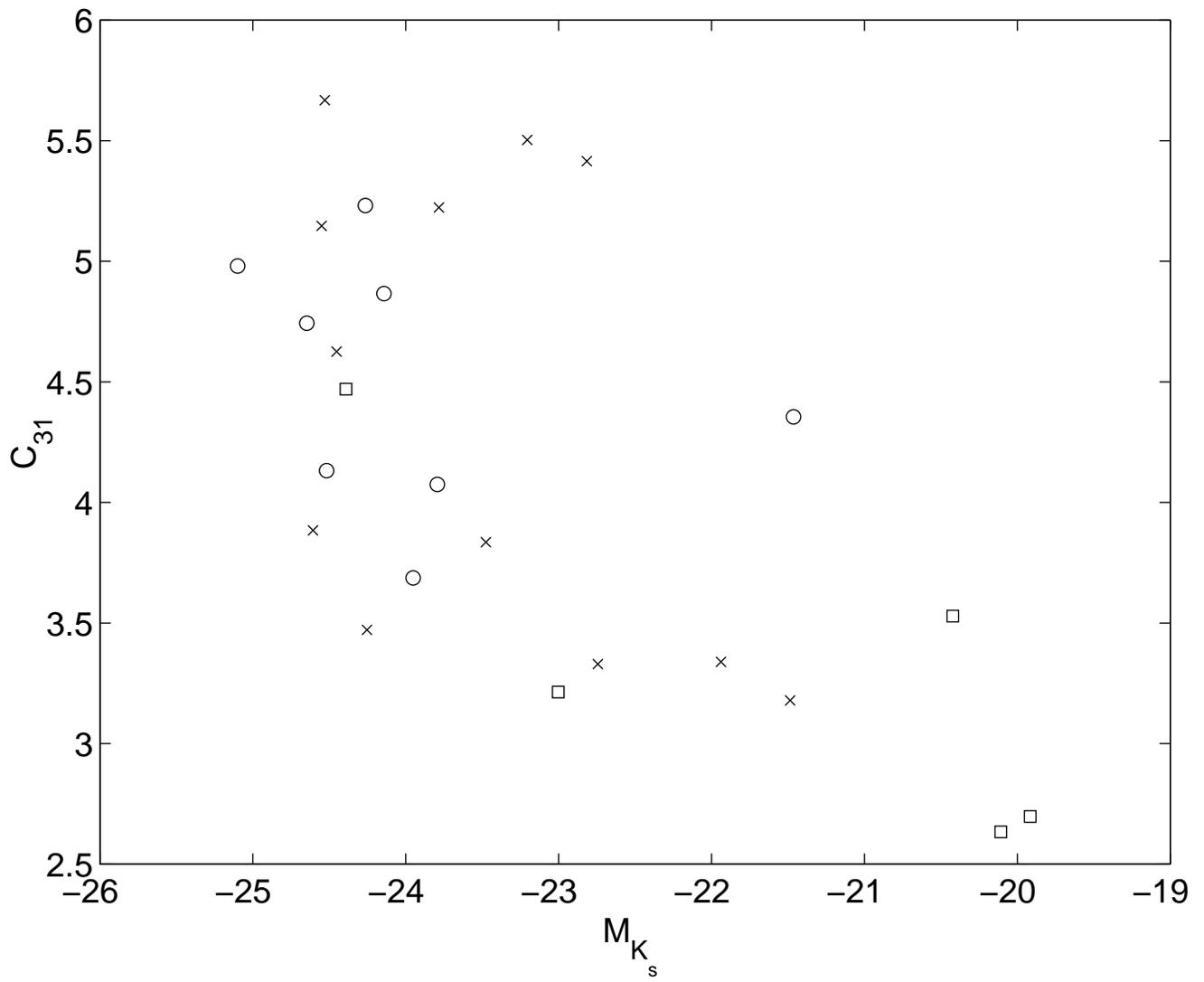
\end{document}